\documentclass[twocolumn]{aastex631}
\usepackage[utf8]{inputenc}
\usepackage{amsmath, todonotes}
\submitjournal{AJ}

\newcommand{\teff}{$T_{\rm eff}$}
\newcommand{\logg}{$\log g$}
\newcommand{\metal}{[Fe/H]$_{\mathrm{m}}$}
\newcommand{\gaia}{\textit{Gaia}}
\newcommand{\spae}{{\tt SPAE}}
\newcommand{\gse}{\gaia-Enceladus}
\newcommand{\ocen}{$\omega$~Cen}
\newcommand{\tess}{{\it TESS}}
\newcommand{\ut}{University of Tampa, 401 W Kennedy Blvd, Tampa, FL 33606, USA}

\begin{document}

\title{Combining Astrometry and Elemental Abundances: The Case of the Candidate Pre-\gaia\ Halo Moving Groups G03-37, G18-39, and G21-22 \footnote{Based on observations collected at the European Southern Observatory under ESO programme 090.B-0605(A)}}

\correspondingauthor{Simon C.~Schuler}
\email{sschuler@ut.edu}

\author[0000-0001-7203-8014]{Simon C.~Schuler}
\affiliation{\ut}

\author[0000-0001-5261-3923]{Jeff J. Andrews}
\affiliation{Center for Interdisciplinary Exploration and Research in Astrophysics (CIERA), 1800 Sherman Ave., Evanston, IL, 60201, USA}

\author{Vincent R.~Clanzy II}
\affiliation{\ut}
\affiliation{Clemson University, 118 Kinard Laboratory, Clemson, SC 29634, USA}
\author{Mohammed Mourabit}
\affiliation{\ut}
\affiliation{The University of Texas at Arlington, 701 S. Nedderman Drive, Arlington, TX 76019, USA}
\author[0000-0003-2481-4546]{Julio Chanam{\'e}}
\affiliation{Pontificia Universidad Cat{\'o}lica de Chile, Av.~Vicu{\~n}a Mackenna 4860, 782-0436 Macul, Santiago, Chile}

\author[0000-0001-7077-3664]{Marcel A.~Ag{\"u}eros}
\affiliation{Columbia University, 550 West 120th Street, New York, NY 10027, USA}

\begin{abstract}
While most moving groups are young and nearby, a small number have been identified in the Galactic halo. Understanding the origin and evolution of these groups is an important piece of reconstructing the formation history of the halo. Here we report on our analysis of three putative halo moving groups: G03-37, G18-39, and G21-22. Based on \gaia\ EDR3 data, the stars associated with each group show some scatter in velocity (e.g., Toomre diagram) and integrals of motion (energy, angular momentum) spaces, counter to expectations of moving-group stars. We choose the best candidate of the three groups, G21-22, for follow-up chemical analysis based on high-resolution spectroscopy of six presumptive members. Using a new Python code that uses a Bayesian method to self-consistently propagate uncertainties from stellar atmosphere solutions in calculating individual abundances and spectral synthesis, we derive the abundances of $\alpha$- (Mg, Si, Ca, Ti), Fe-peak (Cr, Sc, Mn, Fe, Ni), odd-$Z$ (Na, Al, V), and neutron-capture (Ba, Eu) elements for each star. We find that the G21-22 stars are not chemically homogeneous. Based on the kinematic analysis for all three groups and the chemical analysis for G21-22, we conclude the three are not genuine moving groups. The case for G21-22 demonstrates the benefit of combining kinematic and chemical information in identifying conatal populations when either alone may be insufficient. Comparing the integrals of motion and velocities of the six G21-22 stars with those of known structures in the halo, we tentatively associate them with the \gaia-Enceladus accretion event.
\end{abstract}

\section{Introduction} \label{sect:intro}
The advent of imaging and spectroscopic all-sky surveys over the past two decades has enabled great advances in the study of Milky Way stellar populations and led to the realization that our Galaxy is teeming with substructure \citep[e.g.,][]{2020ARA&A..58..205H}. Notable examples include the Sagittarius dwarf galaxy \citep[e.g.,][]{1994Natur.370..194I,1995MNRAS.275..429L,2002ApJ...569..245N,2003ApJ...599.1082M}, the dozens of other dwarf galaxies that surround the Milky Way \citep[e.g.,][]{2005ApJ...626L..85W}, the remnants of the \gaia-Enceladus/Sausage major merger $\approx$10~Gyr ago \citep{2018MNRAS.478..611B, 2018Natur.563...85H}, and other dynamically tagged groups \citep[e.g.,][]{2020ApJ...891...39Y,2020ApJ...901...48N,2021ApJ...907...10L}. Today, thanks to data produced from the ground and from space by e.g., the Dark Energy Survey \citep{2016MNRAS.460.1270D}, the Zwicky Transient Facility \citep{2019PASP..131a8002B}, \gaia\ \citep{2018A&A...616A...1G} and \tess\ \citep{2015JATIS...1a4003R}, we are gaining ever deeper understanding of our Galaxy. 

\begin{deluxetable*}{lcrrcrrcrccrrcrr}
\tablecolumns{16}
\tablewidth{0pt}
\tablecaption{Literature Properties of G03-37, G18-39, and G21-22 \label{tab:silva}}
\tablehead{
	\colhead{}&
	\colhead{}&
	\multicolumn{2}{c}{$U^{\prime}$}&
	\colhead{}&
	\multicolumn{2}{c}{$V^{\prime}$}&
	\colhead{}&
	\multicolumn{2}{c}{$\left(U^{\prime2} + W^{\prime2}\right)^{1/2}$}&
	\colhead{}&
	\multicolumn{2}{c}{[Fe/H]}&
	\colhead{}&
	\multicolumn{2}{c}{$V_{\mathrm{rot}}$}\\
	\colhead{}&
	\colhead{}&
	\multicolumn{2}{c}{(km~s$^{-1}$)}&
	\colhead{}&
	\multicolumn{2}{c}{(km~s$^{-1}$)}&
	\colhead{}&
	\multicolumn{2}{c}{(km~s$^{-1}$)}&
	\colhead{}&
	\multicolumn{2}{c}{}&
	\colhead{}&
	\multicolumn{2}{c}{(km~s$^{-1}$)}\\
	\cline{3-4} \cline{6-7} \cline{9-10} \cline{12-13} \cline{15-16}\\
	\colhead{Groups}&
	\colhead{}&
	\colhead{min}&
	\colhead{max}&
    \colhead{}&
    \colhead{min}&
	\colhead{max}&
	\colhead{}&
	\colhead{min}&
	\colhead{max}&
	\colhead{}&
	\colhead{min}&
	\colhead{max}&
	\colhead{}&
	\colhead{min}&
	\colhead{max}
	}
\startdata
G03-37 &&  -60 &   55 && -335 & -255 &&  42 & 100 && -1.79 & -1.00 && -88 & -19 \\
G18-39 && -250 & -165 && -316 & -225 && 162 & 250 && -1.79 & -1.00 && -88 & -19 \\ 
G21-22 &&  210 &  287 && -316 & -202 && 162 & 278 && -1.79 & -1.00 && -88 & -19 \\
\enddata

\tablecomments{The data are the range of values characteristic of each candidate group given in \citet{2012RMxAA..48..109S}. $U^{\prime}$, $V^{\prime}$, and $W^{\prime}$ velocities were derived using radial velocities and proper motions from multiple literature sources and the matrix equations of \citet{1987AJ.....93..864J}, and corrected for the Sun's velocity $(U,V,W)_{\odot} = (+10.0, +14.9, +7.7)~\mathrm{km~s}^{-1}$. [Fe/H] were determined using new and literature Str{\"o}mgren photometry and the calibration equations of \citet{schuster}, and $V_{\mathrm{rot}} = V^{\prime} + 220~\mathrm{km~s}^{-1}$, where $220~\mathrm{km~s}^{-1}$ was taken as the velocity of the local standard of rest about the Galactic center.}

\end{deluxetable*}

For instance, it is now generally accepted that the Galactic halo is composed of two stellar populations, characterized by the blue and red sequences of the \gaia\ Herztsprung-Russell diagram \citep{2018A&A...616A...1G}. Stars on the blue sequence tend to be relatively metal-poor, more deficient in $\alpha$-elements (O, Mg, S, Si, Ca, Ti), and have kinematics that are consistent with an accreted population. By contrast, stars on the red sequence are relatively more metal-rich, have more enhanced $\alpha$-element abundances, and have kinematics suggesting these stars formed in situ, either in the disk and subsequently dynamically heated to halo-type orbits, or in the halo itself \citep[e.g.,][]{2007Natur.450.1020C, 2010A&A...511L..10N,2011A&A...530A..15N,2012A&A...538A..21S,2012ApJ...757..164R,2018ApJ...863..113H,2015MNRAS.453..758H}.

Identifying halo stars that formed in situ is critical to fully elucidating the evolution of the Galaxy. Here, we examine three halo moving groups from \citet{2012RMxAA..48..109S}, who identified seven new halo moving groups tentatively associated with \ocen\ and the Kapteyn moving group. We note that the Kapteyn moving group had been long hypothesized to be tidal debris from \ocen\ \citep[e.g.][]{2010AJ....139..636W}, but this was proven to be unlikely by means similar to the present work \citep{2015ApJ...808..103N}. \citet{2012RMxAA..48..109S} identified their moving groups using a combination of Str{\"o}gren photometry of 142 metal-poor stars obtained at the Observatorio Astron{\'o}mico Nacional at San Pedro M{\'a}rtir in Baja California, Mexico, and photometry, radial velocities, and proper motions of additional stars from multiple literature sources. Using well established calibrations, they determined distances, metallicities (i.e., [Fe/H]), absolute magnitudes ($M_{\mathrm{V}}$), and Galactic space velocities corrected for the Sun's velocity ($U^{\prime}$, $V^{\prime}$, $W^{\prime}$) for a total of 1695 stars. 

    \begin{figure*}[t!]
    \centerline{\includegraphics[trim=.5cm .55cm .25cm .25cm, clip=True, width=.99\textwidth]{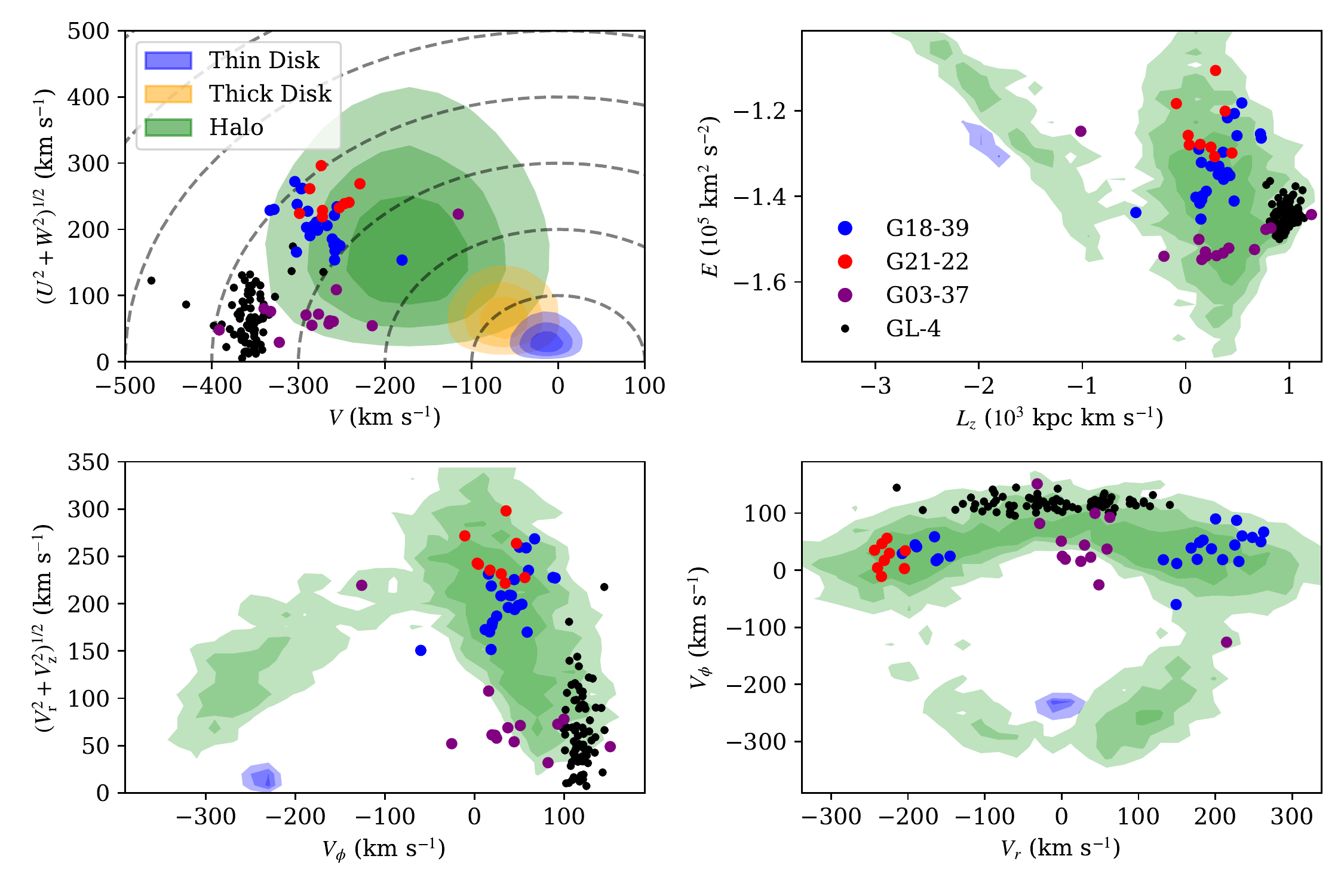}}
    \caption{ We compare the stars in our three  moving groups to stars from the thin disk (blue contours), thick disk (orange contours), halo (green contours), and from the halo substructure GL-4 \citep{2019ApJ...874...74L}, for different kinematic metrics describing their motion around the Galaxy. The top left panel shows a Toomre diagram (the contours in this panel show synthetic stars, while all other panels show stars from \gaia\ DR2 that are kinematically selected as being part of either the thin disk or halo), $\left(U^2 + W^2\right)^{1/2}$ vs.~$V$, comparing the stars' Galactic space velocities. The other panels compare $E$ vs.~$L_{\mathrm{z}}$, the orbital energy against the azimuthal angular momentum (top right), $(V^2_{\mathrm{r}} + V^2_{\mathrm{z}})^{1/2}$ vs.~$V_{\phi}$, the combined radial and longitudinal velocities against the azimuthal velocity around the Galaxy (bottom left), and $V_{\phi}$ vs.~$V_{\mathrm{r}}$,  the azimuthal velocity against the radial velocity. While none of the three \citet{2012RMxAA..48..109S} moving groups is as tightly clustered as are the members of GL-4, the stars in G21-22 do appear relatively clustered, suggesting it is the most likely bona fide moving group of the three. \label{fig:L_z_E}}
    \end{figure*}

We report our first application of a two-pronged approach to validating stellar associations, \gaia\ astrometry and chemical abundance analysis, i.e., chemical tagging. We adopted as a starting point the definition of a moving group from \citep{1958MNRAS.118...65E}, which associates stars based on common space motions. This implies that the stars share a common orbital path about the Galactic center and have done so for at least some amount of time in the past. We further constrained this idea to include the notion that the stars have a shared formation history, i.e., they formed from the same molecular cloud at about the same time. In this scenario, the stars were once members of a now dissociated open cluster or other conatal stellar association and thus, in addition to their shared kinematics, they should have the same age and composition. In addition to this definition, other possible interpretations of moving groups exist, including stars trapped in dynamical resonances arising from the bar or spiral arms \citep[e.g.,][]{2020MNRAS.492.3816A} or accreted stars from dwarf galaxies \citep[e.g.,][]{2018ApJ...863..113H}, both sets of which could share common space motions but not necessarily formation histories.

We started with three putative halo moving groups identified by  \citet{2012RMxAA..48..109S} and for which we had high-resolution spectroscopy: G03-37, G18-39, and G21-22. As is the case with all the groups in \citet{2012RMxAA..48..109S}, the authors named these moving groups after candidate member stars that appear in the Lowell Proper Motions Survey \citep{1959LowOB...4..136G,1961LowOB...5...61G,1975LowOB...8....9G}. In the sections that follow, we describe our use of precise \gaia\ proper motions and parallaxes, as well as \gaia\ radial velocities, to investigate the kinematic properties and dynamical histories of the stars in each group. We then describe our detailed chemical abundance analysis of six stars in G21-22, which we identify as the best candidate to constitute a coherent moving group. Our analysis relied on our new abundance derivation code that uses Bayesian and Markov chain Monte Carlo (MCMC) methods to self-consistently derive stellar parameters, abundances, and uncertainties. We conclude the paper with a discussion on the validity of the G03-38, G18-39, and G21-22 moving groups, and the potential origin of the G21-22 stars.

\section{\gaia\ Astrometry} \label{sec:gaia}
We first analyzed the dynamical characteristics of the stars in the three moving groups from \citet{2012RMxAA..48..109S} on which we focus. A summary of the properties of these groups as presented by \citet{2012RMxAA..48..109S} is given in Table~\ref{tab:silva}. We found that nine G03-38, 15 G18-39, and six G21-22 stars have counterparts in the early third \gaia\ data release \citep[EDR3;][]{2021A&A...649A...1G, 2021A&A...649A...2L} with measured radial velocities. For these stars we have complete six-dimensional phase space information (see Table~\ref{tab:1}), allowing us to fully analyze the kinematics of the stars in these groups. All kinematic quantities have been derived using the Galactic modeling framework of {\tt Astropy v4.0} \citep{astropy:2013,astropy:2018} and the Galactic modeling code, {\tt gala} \citep{2017JOSS....2..388P}, with its built-in Galactic potential {\tt MilkyWayPotential} (see Table~\ref{tab:2}). This model combines the disk model from \citet{2015ApJS..216...29B} with a spherical NFW profile \citep{1997ApJ...490..493N}. 

In the top left panel of Figure~\ref{fig:L_z_E}, we show a Toomre diagram of the stars in each moving group, plotting their combined $(U^2+W^2)^{1/2}$ velocity against $V$. By comparing the positions of the stars in the three \citet{2012RMxAA..48..109S} moving groups to those of the Milky Way thin disk, thick disk, and halo populations defined in \citet{2003A&A...410..527B}, we concluded  that the stars in each of the three groups are indeed consistent with being members of the Milky Way halo.

In the remaining three panels, we compare the stars' integrals of motion, i.e., orbital energy, $E$, against their azimuthal angular momentum, $L_{\mathrm{z}}$ (top right), the azimuthal velocity, $V_{\phi}$, around the Galaxy against the combined radial, $V_{\mathrm{r}}$, and longitudinal, $V_{\mathrm{z}}$, velocities (bottom left), and $V_{\phi}$ against $V_{\mathrm{r}}$ (bottom right). Background contours in each of the three panels represent a random selection of \gaia\ EDR3 stars with measured radial velocities; halo stars (green contours) are defined by having $(U^2 + V^2 + W^2)^{1/2} >$~300~km~s$^{-1}$, while thin disk stars (blue contours) are defined as having $(U^2 + V^2 + W^2)^{1/2} <$~25~km~s$^{-1}$. In all four panels of Figure~\ref{fig:L_z_E}, we also compare the three halo moving groups to GL-4, a halo substructure identified in \gaia\ and LAMOST data by \citet{2019ApJ...874...74L}. 

Regardless of which kinematic parameters are being used, the stars in each of the three \citet{2012RMxAA..48..109S} groups show significant scatter---more than is observed within GL-4---calling into question their designation as a moving group. Of the three groups, G21-22 (red symbols) shows the least scatter.

We also used {\tt gala} and its Galactic potential {\tt MilkyWayPotential} to integrate stellar orbits around the Milky Way. We defined each star's position in phase space using its astrometric parameters \citep[position, proper motion, and parallax;][]{2021A&A...649A...2L} and \gaia\ EDR3 radial velocity \citep[measured by \gaia's Radial Velocity Spectrometer;][]{2018A&A...616A...5C}, then integrated each stars' position backward in time for 200~Myr. 

\begin{figure}[h]
    \begin{center}
    \includegraphics[trim=0.25cm 0.35cm 0.35cm 0.35cm, clip=True,width=0.99\columnwidth,angle=0]{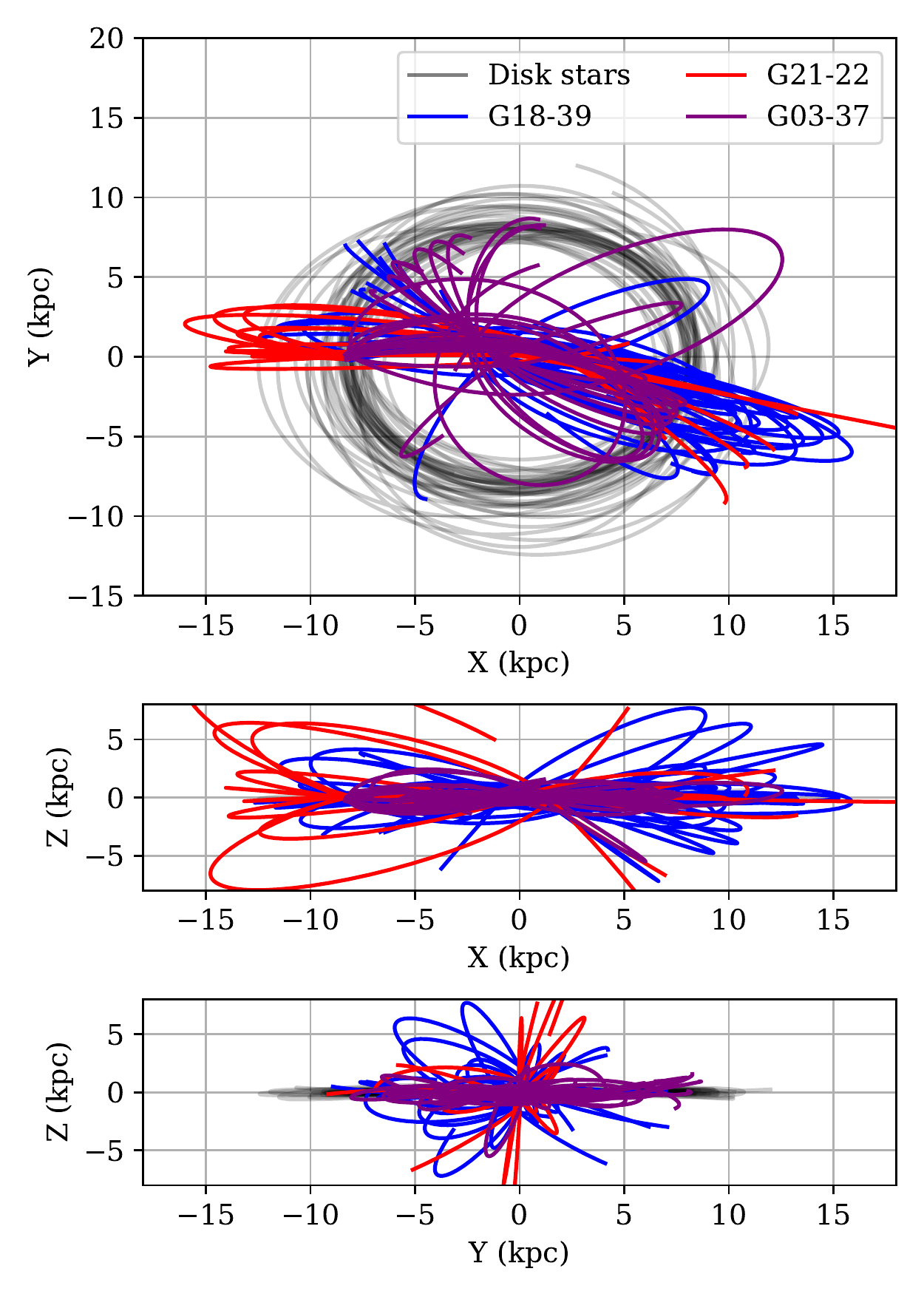}
    \caption{ We integrate the orbit of stars (colored lines) in each of the three stellar groups in a Milky Way potential backwards in time for 200~Myr. Compared to randomly chosen disk stars (black lines) the stars in each of the three stellar groups exhibit halo-like orbits. At the accuracy afforded by \gaia\ EDR3 astrometric precision and radial velocity measurements, none of these three groups appear to have a common origin. \label{fig:orbits}}
    \end{center}
    \end{figure}

Figure~\ref{fig:orbits} shows that the stars in each of these moving groups do not follow kinematically coherent orbits. Even for G21-22, the \citet{2012RMxAA..48..109S} moving group we identified in Figure~\ref{fig:L_z_E} as having the least scatter, the stellar orbits diverge (see the bottom two panels of Figure~\ref{fig:orbits}). This orbital inconsistency strongly suggests that these three moving groups are actually spurious stellar associations. However, uncertainties in the measurement of astrometric parameters, particularly in the parallax, or details of the Galactic model can lead to inaccurate orbits. Therefore, to further constrain the nature of these putative moving groups, we selected the best candidate, G21-22, for a detailed abundance analysis.

    \begin{figure*}[t]
    \centerline{\includegraphics[trim=6cm 3cm 6cm 5cm, clip=True,width=.99\textwidth]{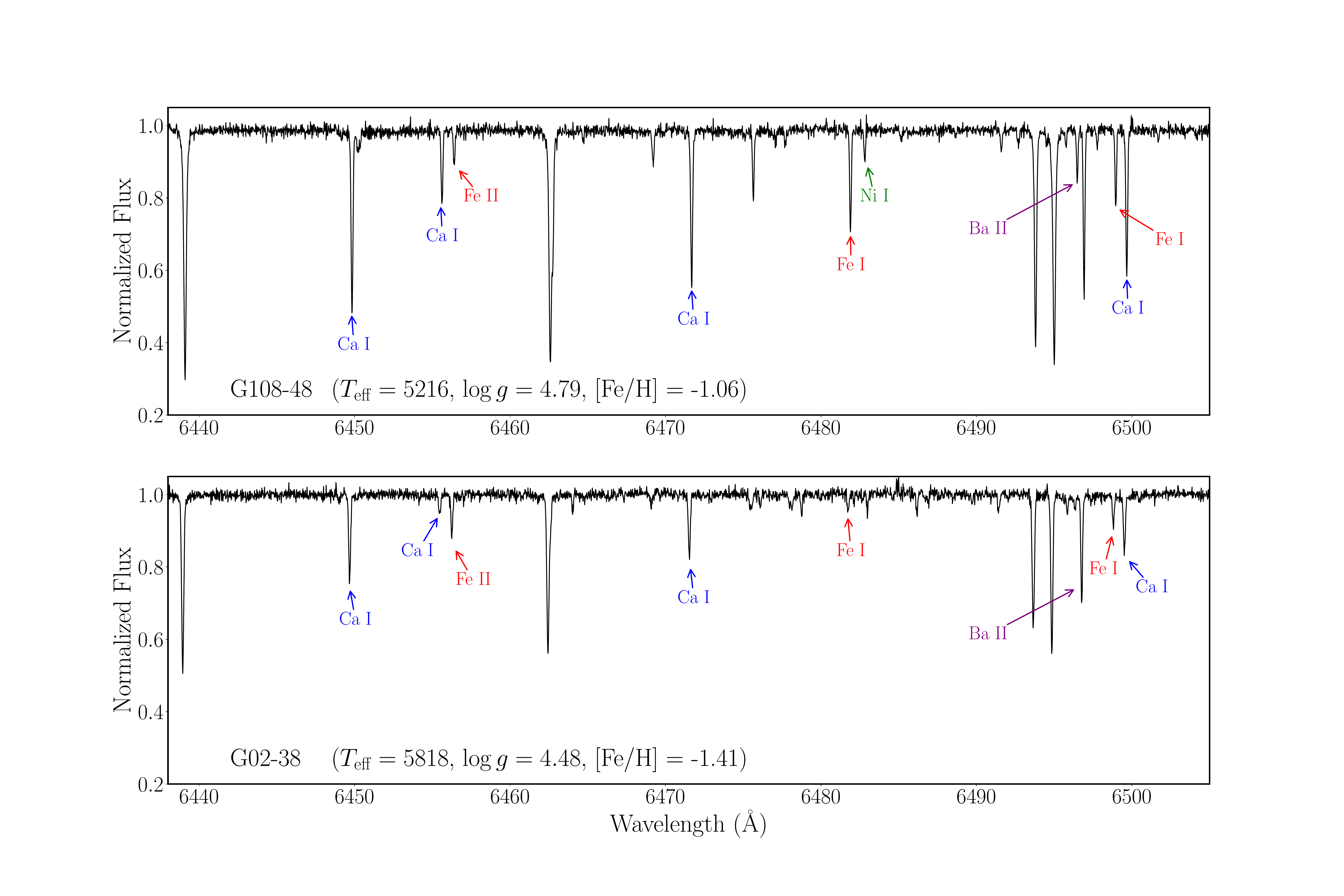}}
    \caption{Sample VLT/UVES spectra of G108-48 (top) and G02-38 (bottom), two putative members of G21-22. Some of the lines that have been measured are indicated. \label{fig:spec}}
    \end{figure*}

\section{Chemical Compositions} \label{sec:comps}
\subsection{Observations and Data Reduction} \label{sub:obs}
We obtained high-resolution spectroscopy of six stars identified by \citet{2012RMxAA..48..109S} as members of G21-22 in service mode with the Ultraviolet and Visual Echelle Spectrograph \citep[UVES;][]{2000SPIE.4008..534D} and the 8.2-m Unit 2 (UT2, Kueyen) telescope of the Very Large Telescope (VLT) at Cerro Paranal, Chile, between the months of 2012 October and 2013 March. A standard dichroic setting (DIC 1) was used to provide a wavelength (incomplete) coverage of 3280 to 6800 {\AA}. The slit width was set to $0.6''$, giving a nominal resolving power of $R\approx65,000$, and the Poisson signal-to-noise (S/N) of the spectra is $\approx$100 near 6500 {\AA}. The data were reduced through the observatory's standard pipeline processing using the REFLEX environment \citep{2013A&A...559A..96F} within the European Southern Observatory Common Pipeline Library and were taken directly from the ESO Science Archive. Representative samples of the spectra are shown in Figure \ref{fig:spec}.

\subsection{Equivalent Widths} \label{subsub:ews}
We derived stellar parameters and abundances of up to 13 elements---Na, Mg, Al, Si, Ca, Sc, Ti, V, Cr, Mn, Fe, Ni, and Ba---via an equivalent width (EW) analysis assuming local thermodynamic equilibrium (LTE) line formation for each star. Starting with the spectral lines analyzed by \citet{2011A&A...535A..31V} and \citet{2019ApJ...874..148S}, we constructed our linelist with those falling in the wavelength coverage of our VLT/UVES spectra. The transition probabilities ($\log gf$) were checked against atomic parameters based on recent accurate laboratory data provided by the Wisconsin atomic physics program \citep[e.g.,][]{2018ApJS..237....8Y,2019ApJS..241...21L}. The laboratory data were accessed via the Python tool {\tt Linemake}\footnote{\url{https://github.com/vmplacco/linemake}. A list of the laboratory studies contributing to this database can be found at \url{http://www.as.utexas.edu/chris/lab.html}.}, developed by Vinicius Placco \citep{2021RNAAS...5...92P}. In all cases of discrepancy, the laboratory transition probabilities were adopted. The linelist is provided in Table~\ref{tab:linelist}. 
    
\begin{deluxetable*}{lcDDDDDDDDD}
\tablecolumns{20}
\tablewidth{0pt}
\tablecaption{Linelist and Equivalent Widths for Putative Members of G21-22 \label{tab:linelist}}
\tablehead{
	\colhead{}&
	\colhead{}&
	\twocolhead{$\lambda$}&
	\twocolhead{$\chi$}&
	\twocolhead{$\log gf$}&
	\multicolumn{12}{c}{Equivalent Widths}\\
	\cline{9-20}\\
	\colhead{Species}&
	\colhead{}&
	\twocolhead{(\AA)}&
	\twocolhead{(eV)}&
	\twocolhead{}&
	\twocolhead{G02-38}&
	\twocolhead{G82-42}&
	\twocolhead{G108-48}&
	\twocolhead{119-64}&
	\twocolhead{G161-73}&
	\twocolhead{HIP~36878}
	}
\decimals
\startdata                                 
\ion{Na}{1}   && 5682.63 & 2.10 & -0.70 &\nodata &\nodata &  19.8  &\nodata &  11.0  &  21.9  \\
              && 5688.21 & 2.10 & -0.45 &\nodata &  24.6  &  33.3  &  10.1  &  19.2  &  31.5  \\
              && 6160.75 & 2.10 & -1.26 &\nodata &\nodata &   6.9  &\nodata &\nodata &\nodata \\
\ion{Mg}{1}   && 5528.41 & 4.34 & -0.62 &\nodata & 169.3  &\nodata &  80.8  & 121.2  &\nodata \\ 
              && 5711.09 & 4.34 & -1.83 &  32.1  &  58.5  &\nodata &  15.8  &  35.3  &  65.7  \\ 
              && 6318.72 & 5.10 & -1.73 &\nodata &\nodata &   9.3  &\nodata &\nodata &\nodata \\  
              && 6319.24 & 5.10 & -1.95 &   7.4  &\nodata &   5.5  &\nodata &\nodata &\nodata \\  
\ion{Al}{1}   && 6696.02 & 3.14 & -1.35 &\nodata &\nodata &   7.2  &\nodata &\nodata &   3.7  \\
\ion{Si}{1}   && 5684.52 & 4.93 & -1.65 &   9.3  &  14.0  &  13.9  &\nodata &  12.0  &\nodata  \\
              && 5701.10 & 4.93 & -2.05 &\nodata &\nodata &\nodata &\nodata &\nodata &   9.2   \\
\enddata

\tablecomments{Table \ref{tab:linelist} is published in its entirety online in the machine-readable format. A portion is shown here for guidance regarding its form and content.}

\end{deluxetable*}

EWs of each line were measured using the Python tool {\tt pyEW}\footnote{\url{https://github.com/madamow/pyew}}, developed by Monika Adam{\'o}w. The code attempts to fit each spectral line from a user-provided linelist with Gaussian, multiple Gaussian (combined), and Voigt profiles and provides the best fit as the one that minimizes the least squares errors in the fit to the observed line. All lines were inspected visually to ensure the goodness of each fit. The EWs of each measured line are included in Table~\ref{tab:linelist}.
        
\subsection{Stellar Parameters, Abundances, and Errors} \label{subsub:spae}
To derive stellar parameters and abundances, we made use of our newly developed Python code, \spae\ (Stellar Parameters, Abundances, and Errors\footnote{\url{https://github.com/simon-schuler/SPAE}}), that uses a state-of-the-art Bayesian method to self-consistently propagate uncertainties from the stellar atmosphere solutions in calculating individual abundances. \spae\ employs the LTE plane-parallel spectral analysis code MOOG\footnote{\url{https://www.as.utexas.edu/~chris/moog.html}} \cite[MOOGSILENT version;][]{1973ApJ...184..839S} and \citet{2011CaJPh..89..417K} model atmosphere grids\footnote{\url{http://kurucz.harvard.edu/grids.html}} to derive the abundances of elements included in a user-provided linelist. We note that in the following and throughout the paper, we distinguish \metal, the metallicity of a star relative to the Sun, and [Fe/H], the Fe abundance of a star relative to the solar Fe abundance\footnote{We use the standard bracket notation to denote abundances relative, in the present case, to solar values, e.g., $\mathrm{[Fe/H]} = \log_{10} [N\mathrm{(Fe)}/N\mathrm{(H)}]_{\star} - \log_{10} [N\mathrm{(Fe)}/N\mathrm{(H)}]_{\sun}$, with $\log_{10} N\mathrm{(H)} = 12.0$.}. The former is used as an input for a model atmosphere; the latter is an output of MOOG.
    
Starting with an (adjustable) initial parameter grid step defined by an effective temperature (\teff), surface gravity (\logg), metallicity (\metal), and microturbulent velocity ($\xi$), the code interpolates the Kurucz grids to produce a model atmosphere, and then passes the model and linelist to MOOG to derive the abundances. The resulting \ion{Fe}{1} and \ion{Fe}{2} abundances are read by \spae\ and used to construct the likelihood function, $\mathcal{L}$, which is defined by the minimization of the differences in the derived [\ion{Fe}{1}/H] and [\ion{Fe}{2}/H] abundances and the model metallicity, [Fe/H]$_{\mathrm{m}}$:

    \begin{equation}
    \ln \mathcal{L}_{\mathrm{Fe\,I}} = -N_{\mathrm{Fe\,I}} \times \ln \sigma_{\mathrm{Fe}} + \sum -\frac{\left([\mathrm{Fe\,I/H}] - [\mathrm{Fe/H}]_{\mathrm{m}}\right)^2}{2\sigma_{\mathrm{Fe}}^2}
    \label{eq:1}
    \end{equation}
and
    \begin{equation}
    \ln \mathcal{L}_{\mathrm{Fe\,II}} = -N_{\mathrm{Fe\,II}} \times \ln \sigma_{\mathrm{Fe}} + \sum -\frac{\left([\mathrm{Fe\,II/H}] - [\mathrm{Fe/H}]_{\mathrm{m}}\right)^2}{2\sigma_{\mathrm{Fe}}^2},
    \label{eq:2}
    \end{equation}

\noindent where $\sigma_{\mathrm{Fe}}$ is the standard deviation of the combined line-by-line abundances of [\ion{Fe}{1}/H] and [\ion{Fe}{2}/H], and $N_{\mathrm{Fe\,I}}$ and $N_{\mathrm{Fe\,II}}$ are the number of \ion{Fe}{1} and \ion{Fe}{2} lines, respectively. The log of the likelihood function is then the sum of the two separate log-likelihoods in Equations \ref{eq:1} and \ref{eq:2}:
            
    \begin{displaymath}
    \ln \mathcal{L} = \ln \mathcal{L}_{\mathrm{Fe\,I}} + \ln \mathcal{L}_{\mathrm{Fe\,II}}.
    \end{displaymath}

\begin{deluxetable*}{lcrcrcrcrcrcr}[t]
\tablecolumns{13}
\tablewidth{0pt}
\tablecaption{Stellar Parameters for the Putative G21-22 Members \label{tab:params}}
\tablehead{
	\colhead{}&
	\colhead{}&
	\colhead{G02-38}&
	\colhead{}&
	\colhead{G82-42}&
	\colhead{}&
	\colhead{G108-48}&
	\colhead{}&
	\colhead{G119-64}&
	\colhead{}&
	\colhead{G161-73}&
	\colhead{}&
	\colhead{HIP~36878}
	}
\startdata
\teff\ (K)          && $5818 \, ^{\,+43}_{-\,42}$ && $5413 \, ^{+\,42}_{-\,43}$   && $5212\,^{+\,46}_{-\,47}$     && $6149 \pm 50$                && $5981\,^{+\,36}_{-\,33}$      && $5333 \; ^{+\,32}_{-\,34}$ \\ 
$\log g$ (cgs)      && $4.48 \pm 0.08$           && $4.61 \pm 0.08$              && $4.77\,^{+\,0.11}_{-\,0.10}$ && $4.06\,^{+\,0.12}_{-\,0.13}$ && $3.97\, ^{+\,0.08}_{-\,0.09}$ && $4.55 \pm 0.05$ \\
$\xi$ (km s$^{-1}$) && $1.27 \pm 0.08$           && $0.65\,^{+\,0.13}_{-\,0.18}$ && $0.78\,^{+\,0.18}_{-\,0.21}$ && $1.54 \pm 0.06$              && $1.45 \pm 0.05$               && $0.49\,^{+\,0.13}_{-\,0.18}$ \\ 
$\mathrm{[Fe/H]}_{\mathrm{m}}$   && $-1.41 \pm 0.03$          && $-1.17 \pm 0.04$             && $-1.07 \pm 0.04$             && $-1.56 \pm 0.04$             && $-1.07 \pm 0.03$              && $-1.13 \pm 0.03$ \\
Iterations\tablenotemark{a} && 26,600 && 30,420 && 31,200 && 28,000 && 30,000 && 34,000 \\
\enddata

\tablenotetext{a}{Number of iterations included in ensemble solution after accounting for burn-in and solutions with low acceptance fractions. Our walkers have typical autocorrelation lengths of $\lesssim 100$, meaning the number of independent samples is somewhat lower than that shown.}

\end{deluxetable*}

\begin{deluxetable*}{rrrrrrrrrrrrr}
\tablecolumns{13}
\tablewidth{0pt}
\tablecaption{Abundances of Putative G21-22 Stars \label{tab:abunds}}
\tablehead{
	\colhead{}&
	\colhead{}&
	\colhead{G02-38}&
	\colhead{}&
	\colhead{G82-42}&
	\colhead{}&
	\colhead{G108-48}&
	\colhead{}&
	\colhead{G119-64}&
	\colhead{}&
	\colhead{G161-73}&
	\colhead{}&
	\colhead{HIP~36878}
	}
\startdata
$\mathrm{[Fe\;I/H]}$  && $-1.41 \; \pm 0.03$ && $-1.17 \; \pm 0.04$ && $-1.07 \; \pm 0.04$ && $-1.56 \; \pm 0.04$ && $-1.07 \; \pm 0.03$ && $-1.13 \; \pm 0.03$ \\
$\mathrm{[Fe\;II/H]}$ && $-1.41 \; \pm 0.04$ && $-1.19 \; \pm 0.05$ && $-1.07 \; \pm 0.05$ && $-1.56 \; \pm 0.05$ && $-1.06 \; \pm 0.04$ && $-1.13 \; \pm 0.03$ \\
$\mathrm{[Na/Fe]}$   && \nodata             && $-0.36 \; \pm 0.05$ && $-0.36 \; \pm 0.05$ && $ 0.00 \; \pm 0.04$ && $-0.24 \; \pm 0.05$ && $-0.20 \; \pm 0.04$ \\
$\mathrm{[Mg/Fe]}$   && $ 0.45 \; \pm 0.04$ && $ 0.18 \; \pm 0.11$ && $-0.23 \; \pm 0.05$ && $ 0.31 \; \pm 0.04$ && $ 0.29 \; \pm 0.08$ && $ 0.39 \; \pm 0.03$ \\
$\mathrm{[Al/Fe]}$   && \nodata             && \nodata             && $-0.24 \; \pm 0.05$ && \nodata             &&  \nodata            && $-0.40 \; \pm 0.03$ \\
$\mathrm{[Si/Fe]}$   && $ 0.35 \; \pm 0.07$ && $ 0.14 \; \pm 0.05$ && $ 0.07 \; \pm 0.04$ && \nodata             && $ 0.19 \; \pm 0.09$ && $-0.05 \; \pm 0.17$ \\
$\mathrm{[Ca/Fe]}$   && $ 0.17 \; \pm 0.05$ && $ 0.18 \; \pm 0.06$ && $ 0.07 \; \pm 0.06$ && $ 0.31 \; \pm 0.06$ && $ 0.21 \; \pm 0.05$ && $ 0.13 \; \pm 0.05$ \\
$\mathrm{[Sc/Fe]}$   && $ 0.13 \; \pm 0.06$ && $ 0.11 \; \pm 0.09$ && $ 0.12 \; \pm 0.08$ && $ 0.12 \; \pm 0.08$ && $ 0.02 \; \pm 0.06$ && $ 0.16 \; \pm 0.06$ \\
$\mathrm{[Ti\;I/Fe]}$  && $ 0.19 \; \pm 0.05$ && $ 0.23 \; \pm 0.06$ && $ 0.20 \; \pm 0.07$ && $ 0.34 \; \pm 0.06$ && $ 0.15 \; \pm 0.06$ && $ 0.18 \; \pm 0.05$ \\
$\mathrm{[Ti\;II/Fe]}$ && $ 0.31 \; \pm 0.05$ && $ 0.17 \; \pm 0.08$ && $ 0.22 \; \pm 0.07$ && $ 0.36 \; \pm 0.07$ && $ 0.19 \; \pm 0.06$ && $ 0.33 \; \pm 0.05$ \\
$\mathrm{ [V/Fe]}$   && \nodata             && $ 0.03 \; \pm 0.10$ && $ 0.11 \; \pm 0.08$ && \nodata             &&  \nodata            && $-0.01 \; \pm 0.05$ \\
$\mathrm{[Cr/Fe]}$   && $-0.17 \; \pm 0.06$ && $ 0.02 \; \pm 0.09$ && $ 0.02 \; \pm 0.07$ && $-0.09 \; \pm 0.09$ && $-0.18 \; \pm 0.05$ && $-0.03 \; \pm 0.05$ \\
$\mathrm{[Mn/Fe]}$   && $-0.41 \; \pm 0.08$ && $-0.41 \; \pm 0.09$ && $-0.29 \; \pm 0.06$ && $-0.44 \; \pm 0.05$ && $-0.33 \; \pm 0.05$ && $-0.34 \; \pm 0.06$ \\
$\mathrm{[Ni/Fe]}$   && $-0.03 \; \pm 0.05$ && $-0.10 \; \pm 0.06$ && $-0.08 \; \pm 0.06$ && $-0.09 \; \pm 0.06$ && $-0.16 \; \pm 0.05$ && $-0.07 \; \pm 0.04$ \\
$\mathrm{[Ba/Fe]}$   && $-0.03 \; \pm 0.07$ && \nodata             && $ 0.23 \; \pm 0.10$ && $-0.04 \; \pm 0.08$ && $ 0.02 \; \pm 0.07$ && $ 0.37 \; \pm 0.06$ \\
\enddata
	
\end{deluxetable*}

The parameter space is sampled using {\tt emcee} \citep{2013PASP..125..306F}, a Python implementation of an affine-invariant ensemble sampler Markov-Chain Monte Carlo (MCMC) algorithm \citep{2010CAMCS...5...65G}. We adopted a four-dimensional-- \teff, \logg, \metal, and $\xi$-- flat prior probability distribution constrained by the available Kurucz model atmosphere grids. The ensemble solution (posterior probability distribution) is generated by initializing 40 walkers in an $N$-ball around Solar values (\teff$=$5777 K, \logg$=$4.44, \metal $=$0.01, $\xi$$=$1.38 km s$^{-1}$), then iterating 1000 steps, resulting in $4 \times 10^4$ separate combinations of parameters that are used to produce model atmospheres that are passed onto MOOG to derive abundances.

    \begin{figure}
    \begin{center}
    \includegraphics[width=0.99\columnwidth,angle=0]{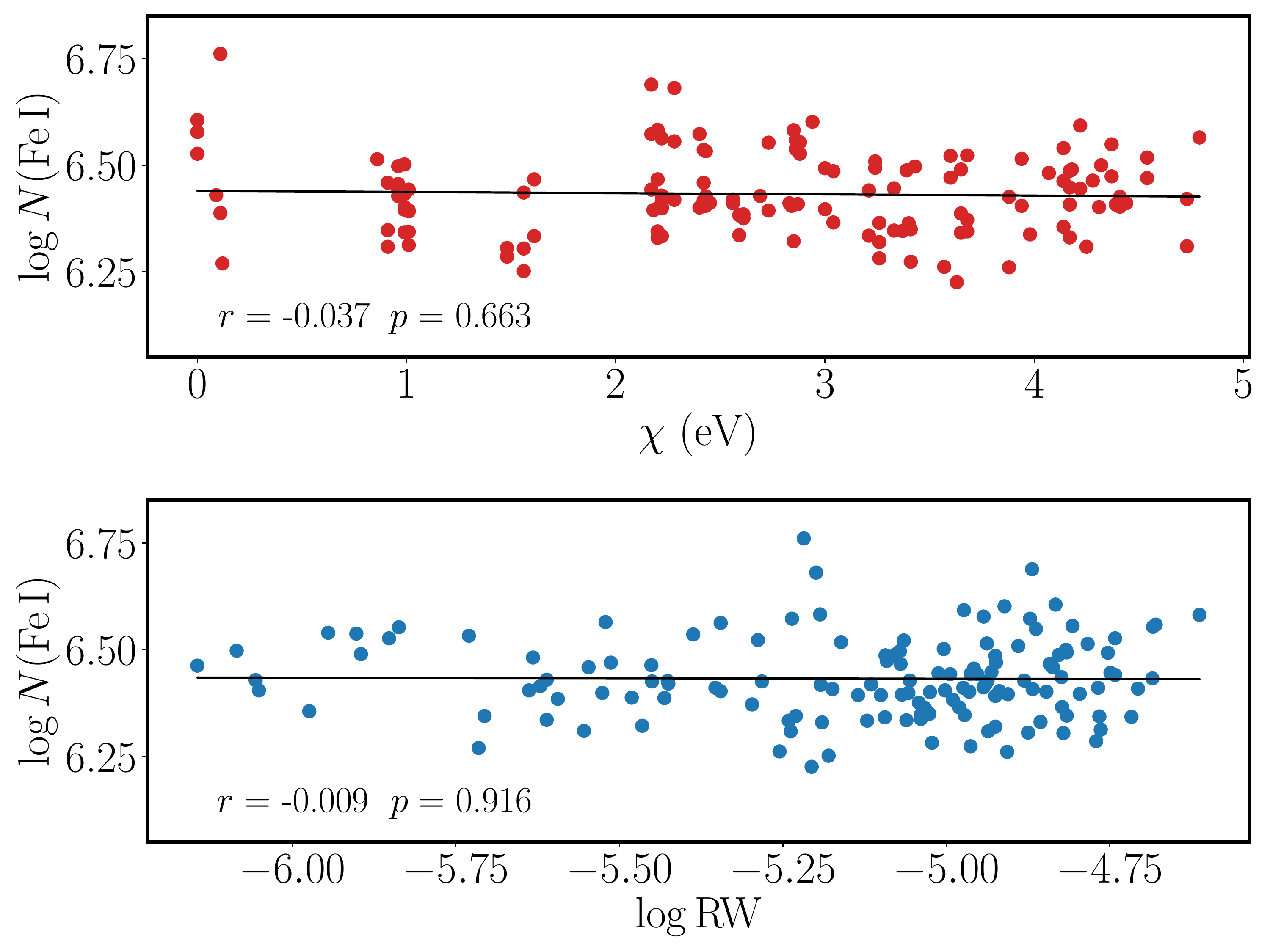}
    \caption{\ion{Fe}{1} abundance versus excitation potential ($\chi$; top) and reduced equivalent width ($\log \mathrm{RW}$; bottom) for the median solution for G161-73. The black line is the least squares fit to the data, with the $r$-value (correlation coefficient) and $p$-value (Wald test with t-distribution) shown in the lower left.\label{fig:median_bal}}
    \end{center}
    \end{figure}

The derived \teff, \logg, \metal, $\xi$, elemental abundances, and their uncertainties are taken as the median and $1\sigma$ confidence intervals of the ensemble solutions, after accounting for burn-in (iterations to convergence, typically 200-300) and any walkers with low acceptance fractions ($<$$0.40$). We calculated autocorrelation lengths with built-in functions in {\tt emcee}, and they were found to be $\lesssim 100$ for each star. Note that \metal\ is determined by considering both \ion{Fe}{1} and \ion{Fe}{2}, as defined by $\mathcal{L}$. The parameters, along with the final number of iterations included in the ensemble solutions, are given for each star in Table~\ref{tab:params}. 

In addition to the stellar parameters and metallicity, the abundances of the other elements measured are derived with each iteration of the MCMC, resulting in ensemble solutions for each element. In Table~\ref{tab:abunds}, the medians and $1\sigma$ confidence intervals of the ensemble solutions for each element are shown, with the abundances given relative to the solar values of \citet{2009ARA&A..47..481A} \footnote{During the preparation of this manuscript, the solar abundances of \citet{2009ARA&A..47..481A} were updated by \citet{2021arXiv210501661A}. With the new solar values, we estimate the [X/Fe] abundances of the G21-22 stars would change by about $\pm 0.02$~dex, except for Ba, the change for which would be about $-0.10$ dex due to the large increase in the solar Ba abundance in \citet{2021arXiv210501661A} relative to \citet{2009ARA&A..47..481A}.}.

The implementation in \spae\ of the likelihood function $\mathcal{L}$ is a departure from the typical method used to spectroscopically derive stellar parameters and abundances. The more typical method, excitation and ionization balance of \ion{Fe}{1} and \ion{Fe}{2}, entails requiring the line-by-line \ion{Fe}{1} abundances to be independent of excitation potential ($\chi$) and EW, and requiring the mean [Fe/H] abundances as derived from \ion{Fe}{1} and \ion{Fe}{2} lines to be the same. The uncertainties are then typically determined by altering the stellar parameters to achieve $1\sigma$ deviations in the correlations between $\chi$, EW, and the \ion{Fe}{1} abundances \citep[e.g.,][]{2011ApJ...732...55S}.

While ionization balance is intrinsic to $\mathcal{L}$, excitation balance is not. Therefore, the ability to check the excitation balance of the median solution, as well as identifying any solution(s) in the ensemble that achieve excitation balance, has been built into \spae. This allows the user to verify that the solution(s) obtained using the typical method is contained in the posterior distribution and determine its (their) statistical significance relative to the distribution.

    \begin{figure}
    \begin{center}
    \includegraphics[width=0.99\columnwidth,angle=0]{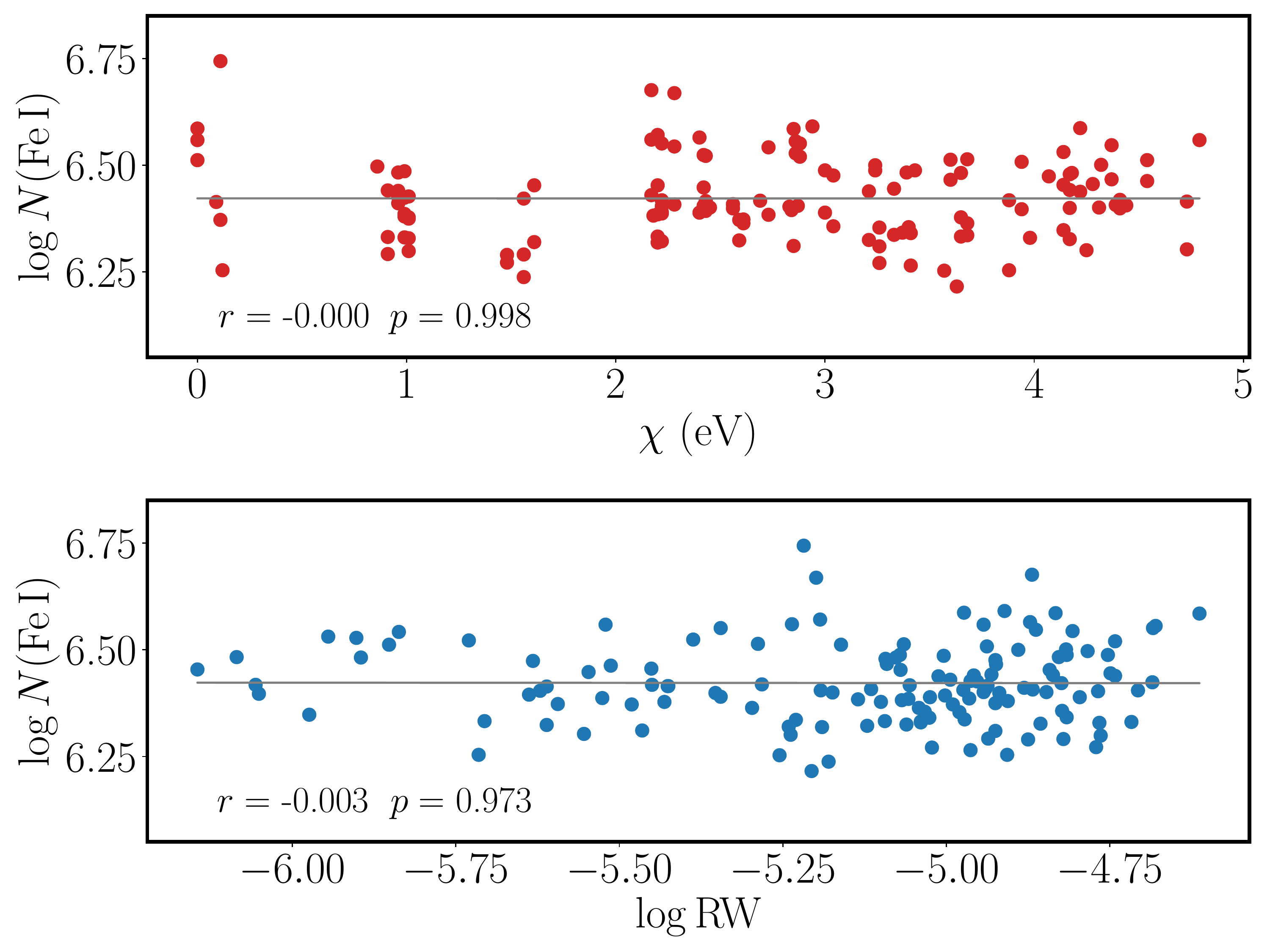}
    \caption{\ion{Fe}{1} abundance versus excitation potential (top) and reduced equivalent width (bottom) for the solution satisfying excitation balance at the highest confidence level for G161-73. The black line is the least squares fit to the data, with the $r$-value and $p$-value shown in the lower left.\label{fig:exc_bal}}
    \end{center}
    \end{figure}

As an example, in Figure~\ref{fig:median_bal} we show the plots of $\log_{10} N$(\ion{Fe}{1}) as a function of $\chi$ and reduced equivalent width ($\log_{10} \mathrm{RW}$) for the median solution for G161-73. A linear least squares fit is applied to each relation. For $\chi$, the $r$-value (correlation coefficient) is $r = -0.037$ and the corresponding $p$-value (Wald test with t-distribution) is $p = 0.663$, and for $\log_{10} \mathrm{RW}$, $r = -0.009$ and $p = 0.916$. While the statistical significance of $r$ is marginal for $\log_{10} \mathrm{RW}$, the $p$-value for $\chi$ indicates that the data are not consistent with the null hypothesis of having a slope of zero and thus formally do not satisfy excitation balance, the criteria for which we have adopted to be $\lvert r \rvert < 0.01$ and $p > 0.90$. However, of the 30,000 iterations included in the ensemble solution for G161-73, 17 of them do satisfy excitation balance. In Figure~\ref{fig:exc_bal} we show the correlation plots for the solution satisfying excitation balance at the highest confidence level for G161-73. With $r = -0.000$ and $p = 0.998$ for $\chi$ and $r = -0.003$ and $p = 0.973$ for $\log_{10} \mathrm{RW}$, this solution is consistent with both relations having a slope of zero. The differences in \teff, \logg, \metal, and $\xi$ between this solution and the median solution are $-9$~K, $-0.07$~dex, $-0.01$~dex, and 0.02~km~$s^{-1}$, respectively.
 
\begin{figure}
    \begin{center}
    \centerline{\includegraphics[trim=.2cm .25cm .5cm .7cm,width=0.99\columnwidth,angle=0]{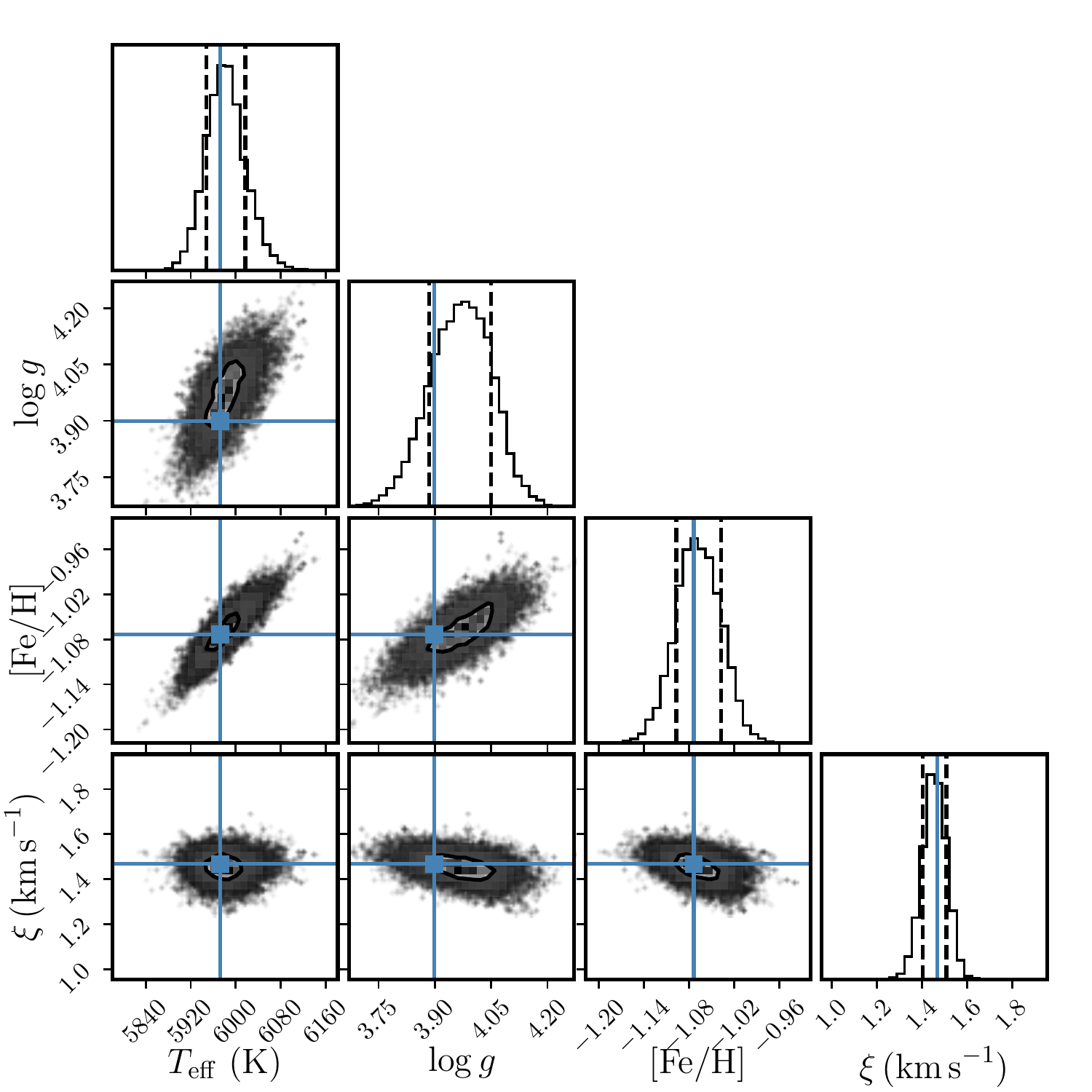}}
    \caption{Stellar parameters of the solution satisfying excitation balance at the highest confidence level for G161-73. Lines indicating the values of the parameters are shown overlaid the ensemble solution (posterior distribution) for the star. The $1\sigma$ confidence intervals are shown as the contours in the scatter plots and vertical dashed lines in the histograms. \label{fig:corner}}
    \end{center}
    \end{figure}

\begin{figure}
    \begin{center}
    \centerline{\includegraphics[trim=.4cm .25cm -0.1cm .4cm, width=0.99\columnwidth,angle=0]{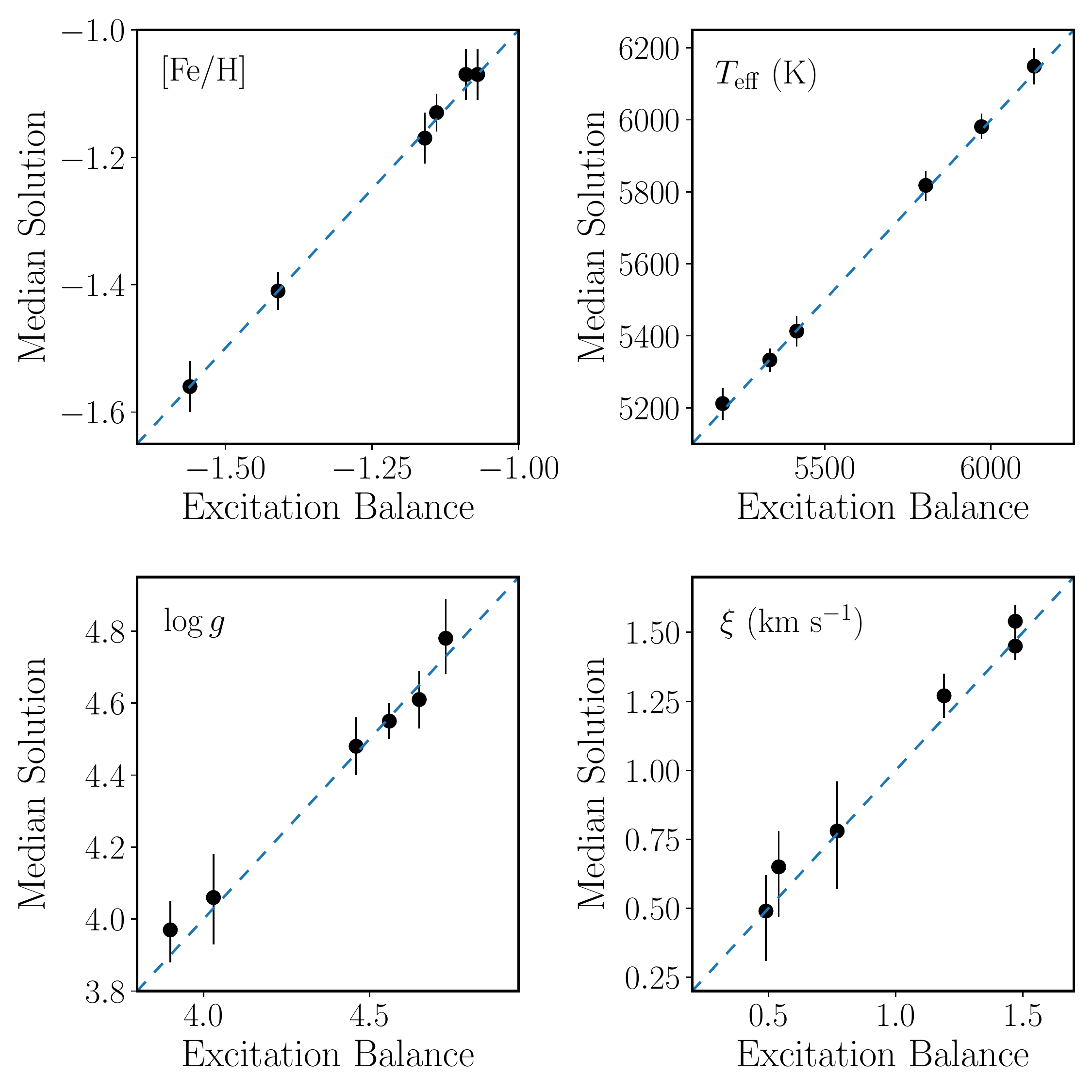}}
    \caption{Stellar parameters of the median solution of the ensemble solutions versus stellar parameters of the solution satisfying excitation balance at the highest confidence level for each star. The dotted line is the line of equality, and errorbars represent the $1\sigma$ confidence intervals of the ensemble solutions. The plots demonstrate that the parameters of the solution satisfying excitation balance is within one standard deviation of the median solution for each. \label{fig:compare}}
    \end{center}
    \end{figure}

The position of this solution is indicated by the blue lines in Figure~\ref{fig:corner}, which summarizes the ensemble solution for the stellar parameters of G161-72 as determined by \spae. All four parameters are within $1\sigma$ of those of the median solution, indicating that \spae\ is able to converge on the correct solution for this star. A direct comparison of the \spae\ solutions to the solutions satisfying excitation balance at the highest confidence level for each star is shown in Figure~\ref{fig:compare}.
    
The mean of the stellar parameters of the solutions satisfying excitation balance for each star is provided in Table~\ref{tab:bal_params}; standard deviations and the number of solutions are also given. The standard deviations demonstrate that although numerous solutions satisfy excitation balance for each star, those solutions occupy a tight parameter space, as would be expected. A major advantage of \spae\ is that these solutions are contained in the generated ensemble solution and therefore taken into account in the final stellar parameter and abundance uncertainties. Indeed, the $1\sigma$ confidence intervals of the ensemble solutions are a true measure of the uncertainty in the parameters and abundances as derived with our data.
        
\begin{deluxetable*}{lcrcrcrcrcrcr}
\tablecolumns{13}
\tablewidth{0pt}
\tablecaption{Excitation Balance Stellar Parameters of Putative G21-22 Stars\tablenotemark{a} \label{tab:bal_params}}
\tablehead{
	\colhead{}&
	\colhead{}&
	\colhead{G02-38}&
	\colhead{}&
	\colhead{G82-42}&
	\colhead{}&
	\colhead{G108-48}&
	\colhead{}&
	\colhead{G119-64}&
	\colhead{}&
	\colhead{G161-73}&
	\colhead{}&
	\colhead{HIP~36878}
	}
\startdata
\teff\ (K)          && $5804.0 \pm   1.8$ && $5415.4 \pm   3.8$ && $5197.4 \pm    3.9$ && $6130.3 \pm   1.3$ && $5970.4 \pm   2.4$ && $5333.2 \pm   3.9$ \\ 
$\log g$ (cgs)      && $4.467  \pm 0.009$ && $4.652  \pm 0.007$ && $4.728  \pm  0.005$ && $4.029  \pm 0.008$ && $3.903  \pm 0.005$ && $4.557  \pm 0.003$ \\
$\xi$ (km s$^{-1}$) && $1.187  \pm 0.006$ && $0.552  \pm 0.017$ && $0.794  \pm  0.016$ && $1.469  \pm 0.005$ && $1.469  \pm 0.004$ && $0.487  \pm 0.023$ \\ 
$\mathrm{[Fe/H]}_{\mathrm{m}}$   && $-1.402 \pm 0.006$ && $-1.161 \pm 0.010$ && $-1.075 \pm  0.010$ && $-1.560 \pm 0.007$ && $-1.077 \pm 0.007$ && $-1.129 \pm 0.008$ \\
Solutions\tablenotemark{b} && 20 && 10 && 14 && 7 && 17 && 24 \\
\enddata

\tablenotetext{a}{The stellar parameters included herein are the means of parameters of the solutions satisfying excitation balance as defined in the text. The quoted uncertainties are the standard deviations of the means.}
\tablenotetext{b}{Number of solutions satisfying excitation balance.}

\end{deluxetable*}

    \begin{figure*}
    \includegraphics[width=0.99\textwidth, angle=0]{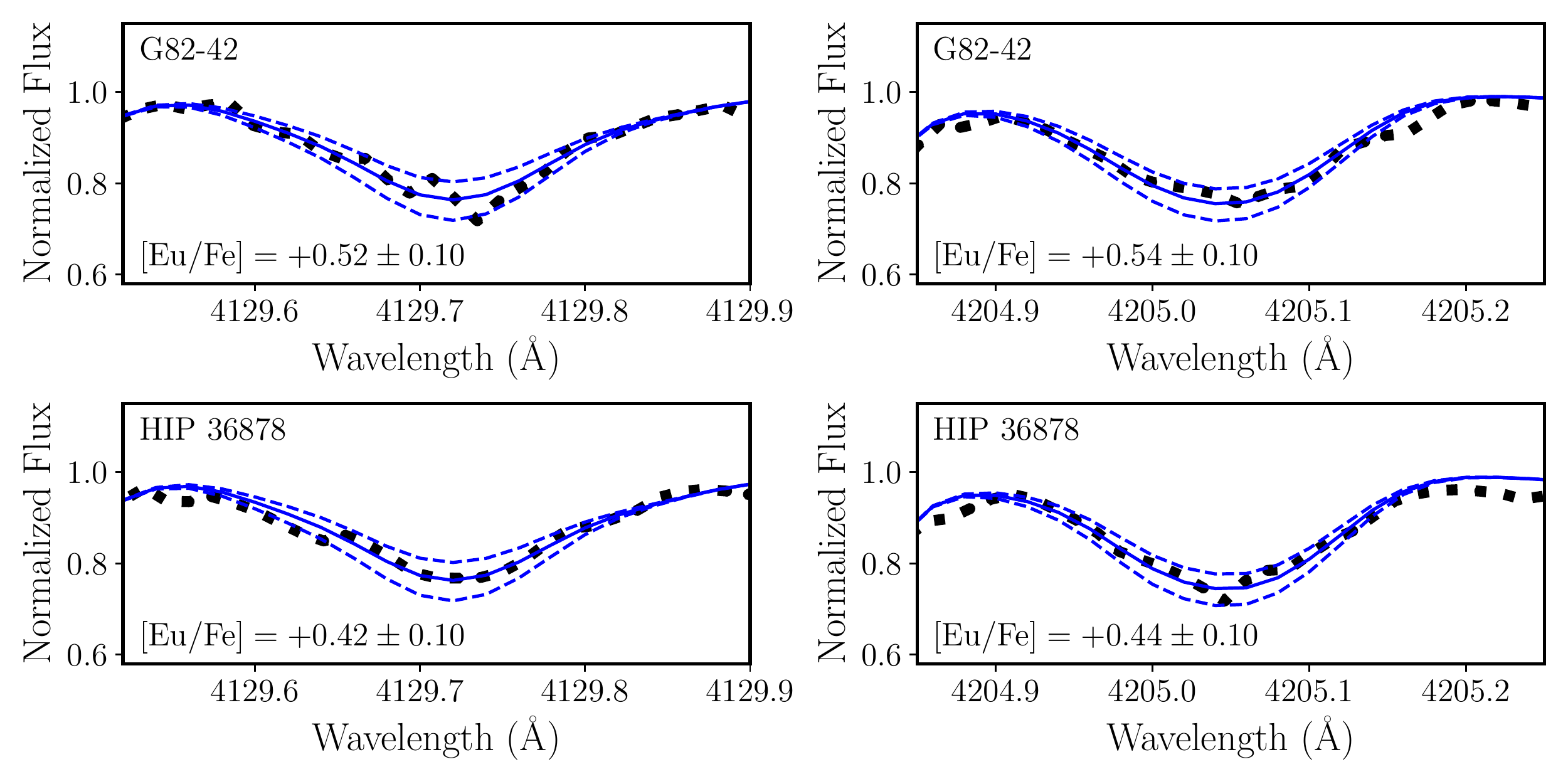}
    \caption{Sample synthesis fits to the \ion{Eu}{2} lines at 4129.7~{\AA} (left) and 4205.0~{\AA} (right) for two stars, G82-42 (top) and HIP~36878 (bottom). Observed spectra are shown as dotted lines, best fits to the observed spectra as solid lines, and $\pm 0.10 \; \mathrm{dex}$ the best fits as dashed lines. \label{fig:eu}}
    \end{figure*}

\subsection{Spectral Synthesis: Europium} \label{sub:syn}
While we used \spae\ to analyze the majority of the elements we discuss in this work, we took a separate approach to deriving the abundances of the predominately $r$-process element Eu. We derived Eu abundances using spectrum synthesis of the \ion{Eu}{2} lines at 4129.7 and 4205.0 {\AA}. Other lines typically used for Eu abundances, such as $\lambda 3724$, $\lambda 3819$, $\lambda 6437$, and $\lambda 6645$, were not detected in our spectra. Synthetic spectra were fit to the observed spectra using the \textit{synth} driver of MOOG, along with a stellar model characterized by the parameters and metallicities from \spae. Linelists for the 4205 and 4130 {\AA} regions were assembled using {\tt Linemake}, which includes the atomic data for the \ion{Eu}{2} lines from \citet{2001ApJ...563.1075L}; we adopted isotopic ratios of 47.8\% and 52.2\% for $^{152}\mathrm{Eu}$ and $^{153}\mathrm{Eu}$, respectively, also from \citet{2001ApJ...563.1075L}.

    \begin{figure*}
    \centerline{\includegraphics[trim=2.2cm 1.5cm 2.75cm 2.75cm, clip=True, width=0.99\textwidth,angle=0]{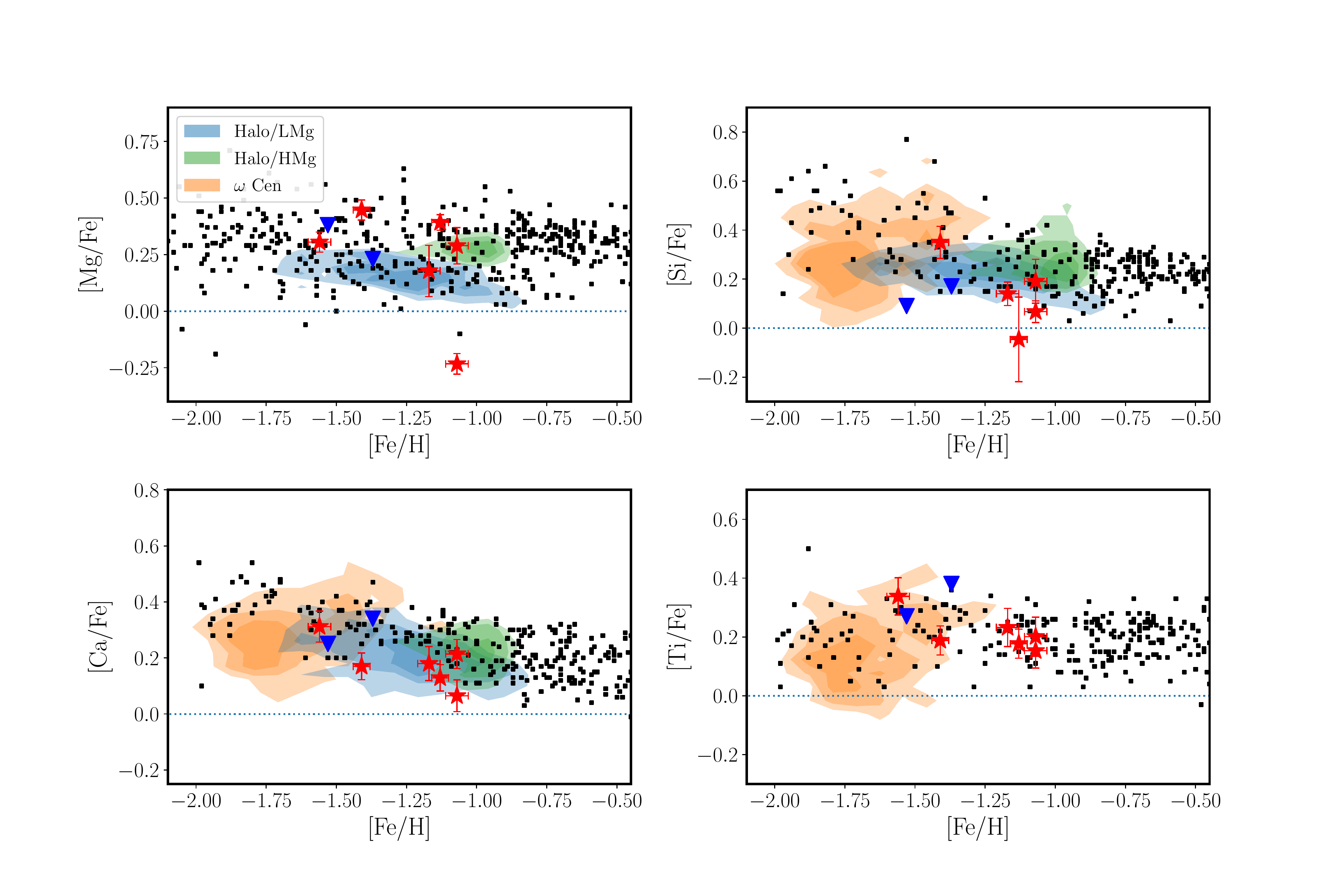}}
    \caption{Abundances of $\alpha$-elements as a function of metallicity. G21-22 abundances are given as red stars with error bars representing the final adopted uncertainties listed in Table~\ref{tab:abunds}. The blue and green contours represent the abundances of halo stars designated as likely to have been accreted by the Milky Way (LMg) and to have formed in situ (HMg), respectively, from \citet{2018ApJ...852...49H}. The orange contour represents the abundances of the large sample of \ocen\ stars from \citet{2010ApJ...722.1373J}. Blue inverted triangles are abundances of the two stars classified as part of \gse\ from \citet{2020MNRAS.497.1236M}. Black squares are general halo star abundances from \citet{2004AJ....128.1177V}, \citet{2006MNRAS.367.1329R}, \citet{2010A&A...511L..10N}, \citet{2012ApJ...753...64I}, and \citet{2014AJ....147..136R}. \label{fig:alpha}}
    \end{figure*}

The linelist was further adjusted by changing slightly the transition probabilities and wavelengths of a handful of lines in the $\lambda 4130$ and $\lambda 4205$ regions to fit a high-quality solar spectrum \citep{2015ApJ...815....5S} obtained with the W.M. Keck Observatory and High Resolution Echelle Spectrometer \citep[HIRES;][]{1994SPIE.2198..362V}. None of the atomic parameters of the Eu lines were altered during this process.

Examples of the synthetic fits to the Eu lines are shown in Figure~\ref{fig:eu}. These lines are found in the bluer wavelength regions of our spectra and thus are characterized by lower S/N ratio compared to the redder regions where the rest of the metal lines are measured. The effect of the lower S/N on the Eu lines themselves can be seen in Figure~\ref{fig:eu}. Nonetheless, we measured consistent [Eu/Fe] abundances from each line for each star. The derived [Eu/Fe] abundances are given in Table~\ref{tab:europium}. We adopted a conservative abundance uncertainty of $\pm 0.10 \, \mathrm{dex}$ for each star.
        
\begin{deluxetable}{lcccccc}
\tablecolumns{7}
\tablewidth{0pt}
\tablecaption{Europium Abundances \label{tab:europium}}
\tablehead{
	\colhead{Star}&
	\colhead{}&
	\colhead{$\mathrm{[Eu/Fe]}_{4129}$}&
	\colhead{}&
	\colhead{$\mathrm{[Eu/Fe]}_{4205}$}&
	\colhead{}&
	\colhead{$<\mathrm{[Eu/Fe]}>$}
	}
\startdata
G02-38    && +0.53 && +0.55 && +0.54 \\
G82-42    && +0.52 && +0.54 && +0.53 \\
G108-48   && +0.57 && +0.60 && +0.59 \\
G119-64   && +0.50 && +0.50 && +0.50 \\
G161-73   && +0.35 && +0.33 && +0.34 \\
HIP 36878 && +0.42 && +0.44 && +0.43 \\
\enddata
	
\end{deluxetable}

\section{Discussion} \label{sec:disc}
\citet{2012RMxAA..48..109S} designated G03-37, G18-39, and G21-22 as moving groups based on stellar density contours in kinematic-metallicity ([Fe/H] vs.~$V_{\mathrm{rot}}$) and kinematic (Bottlinger and Toomre) diagrams constructed using Str{\"o}mgren photometry, radial velocities, and proper motions, mostly obtained from disparate literature sources. We have reproduced this analysis with complete six-dimensional phase space information from \gaia\ EDR3; the related kinematic and integrals of motion diagrams are shown in Figure~\ref{fig:L_z_E}. Such diagrams are helpful in identifying smaller stellar associations \citep[e.g.,][]{2020ApJ...891...39Y,2021ApJ...907...10L}, as well as larger structures \citep[e.g.,][]{2020ARA&A..58..205H}, because related stars will generally cluster together in the metallicity and kinematic phase space.

    \begin{figure*}
    \centerline{\includegraphics[trim=2.5cm 1.5cm 2.75cm 2.75cm, clip=True, width=0.99\textwidth,angle=0]{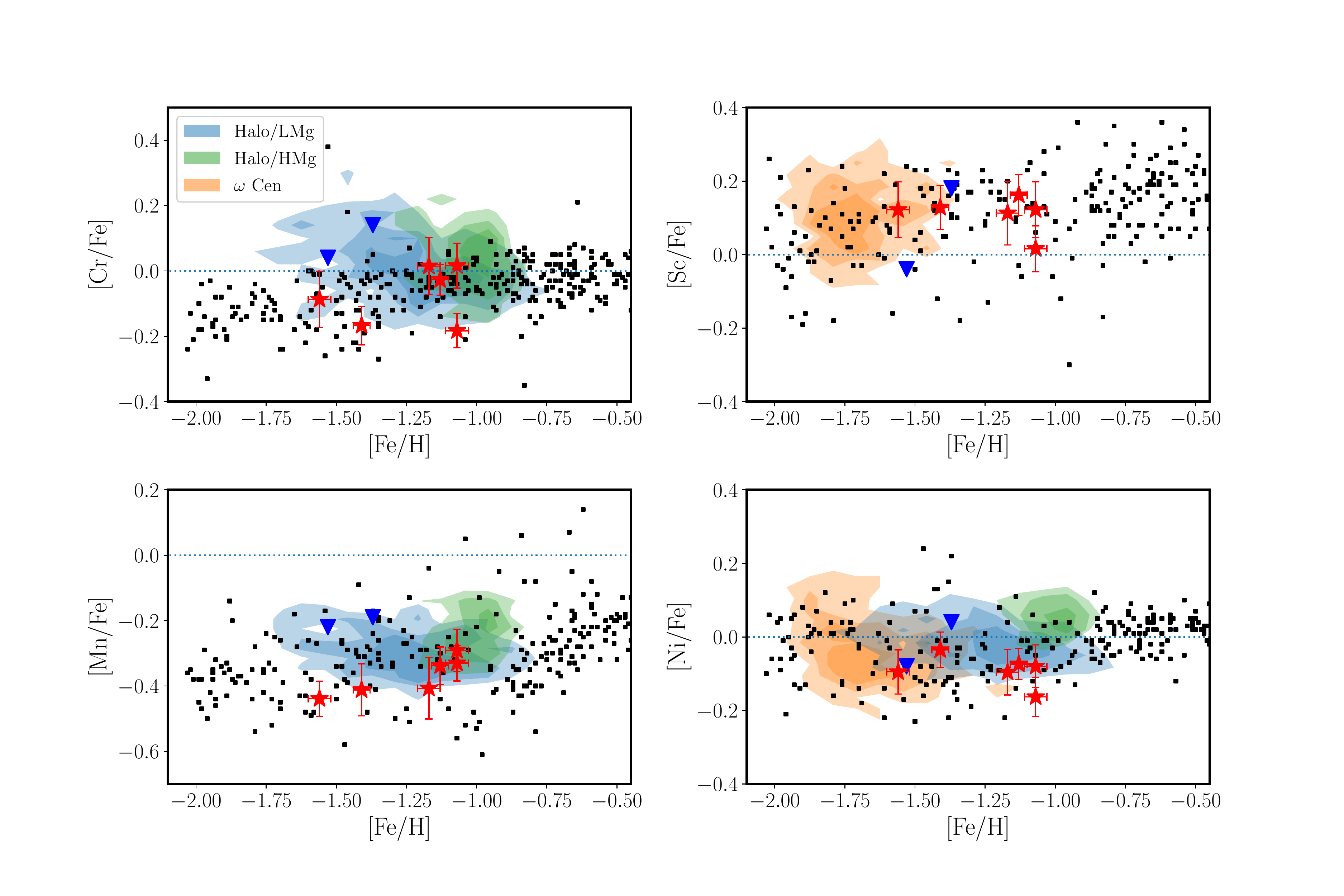}}
    \caption{Abundances of Fe-Peak elements as a function of metallicity. Symbols and sources are the same as given in Figure~\ref{fig:alpha}. Additional halo star abundances from \citet{2013ApJ...771...67I} are also included. \label{fig:fePeak}}
    \end{figure*}

As an example, in Figure~\ref{fig:L_z_E} we include 75 stars of the GL-4 halo substructure identified by \citet{2019ApJ...874...74L}. These stars form a tight group in the integrals of motion and occupy narrow regions in the velocity spaces. If G03-37, G18-39, and G21-22 were true moving groups, a similar degree of clustering would be expected. However, this is not observed for G03-37 nor G18-39, both of which demonstrate larger scatter than GL-4 in the four diagnostics of Figure~\ref{fig:L_z_E}. The G21-22 stars also do not cluster tightly together, although the scatter is reduced compared to the other groups.

With the assumption that moving group stars should not only share common space motions but also compositions, detailed chemical abundance analysis is a powerful tool to complement a dynamical analysis of putative moving group members. Each alone may not be definitive in identifying bona fide members. With the \gaia-based dynamical analysis shown in Figure~\ref{fig:L_z_E},  as well as the orbital analysis shown in Figure~\ref{fig:orbits}, seemingly ruling out G03-37 and G18-39 as moving groups but leaving open the possibility for G21-22, we derived detailed abundances for the six stars in that group.

    \begin{figure*}
    \centerline{\includegraphics[trim=2.2cm 1.5cm 2.75cm 2.75cm, clip=True, width=0.99\textwidth,angle=0]{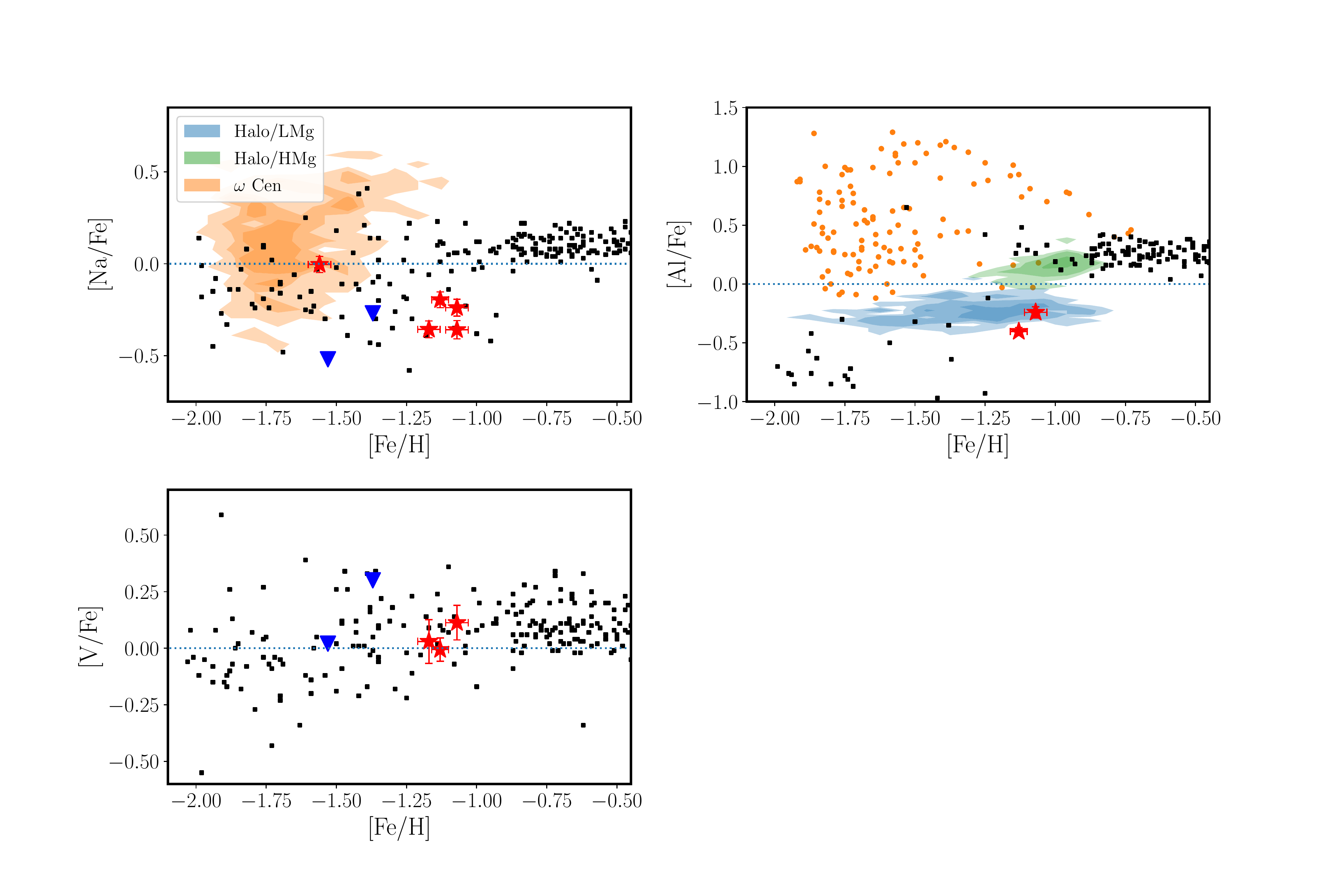}}
    \caption{Abundances of the odd-$Z$ elements Na, Al, and V as a function of metallicity. Symbols and sources are the same as given in Figure~\ref{fig:alpha}, except for Al, where the orange contour for \ocen\ stars has been replaced with orange circles due to the paucity of these stars with Al abundances. Additional halo star abundances from \citet{2013ApJ...771...67I} are also included. \label{fig:oddZ}}
    \end{figure*}

In Figures~\ref{fig:alpha}, \ref{fig:fePeak}, \ref{fig:oddZ}, and \ref{fig:baeu}, we plot the abundances of $\alpha$- (\ion{Mg}{1}, \ion{Si}{1}, \ion{Ca}{1}, and \ion{Ti}{1}), Fe-Peak (\ion{Cr}{1}, \ion{Sc}{2}, \ion{Mn}{1}, and \ion{Ni}{1}), odd-$Z$ (\ion{Na}{1}, \ion{Al}{1}, and \ion{V}{1}), and $n$-capture (\ion{Ba}{2} and \ion{Eu}{2}) elements, respectively, versus [Fe/H] for the six G21-22 stars and other stars in the Galactic halo. Focusing on the G21-22 stars only, it is clear that the six do not compose a chemically homogeneous group. Four stars (G82-42, G108-48, G161-73, and HIP~36878) are characterized by higher metallicities,  $\mathrm{[Fe/H]} \approx -1.1$, while the other two are more metal poor at $\mathrm{[Fe/H]} \approx -1.4$ (G02-38) and $\approx -1.6$ (G119-64). For the other elements, their abundances relative to Fe demonstrate mild scatter among the six stars, but nothing unusual is apparent, with the possible exception of the Mg abundance of G108-48. For the metal-rich group, some abundances are consistent within the $1\sigma$ confidence intervals among the stars (Si, Ti, V, Mn, Ni) while some are not (Na, Mg, Ca, Sc, Cr, Ba, Eu).

Taken together, the kinematic metrics and detailed compositions of the G21-22 stars strongly suggest that this collection of stars does not constitute a moving group of stars with a common origin. Given the grouping of four stars in kinematic phase space (Figure~\ref{fig:L_z_E}) and the grouping of four at a metallicity $\mathrm{[Fe/H]} \approx -1.1$, it is possible that a subset of the six stars are members of a moving group. However, the stars in each group are not the same. One of the kinematically similar stars (G02-38) is in the metal-poor group, and the star (G108-48) with the most dissimilar kinematic properties is in the metal-rich group.

\subsection{On the Origins of the G21-22 Stars} \label{sub:origin}
By comparing the kinematic properties of the G21-22 stars to the models of \citet{2005MNRAS.359...93M}, \citet{2012RMxAA..48..109S} suggest that the stars may be related to the tidal debris of the long hypothesized \ocen\ accretion event \citep{1993ASPC...48..608F} or some other accreted dwarf galaxy. \citet{2019IAUS..344..134S} offer an alternate explanation, implying the stars formed in situ in the Galaxy and are now trapped in orbital resonances created by the Galactic bar \citep[see also][]{2020MNRAS.492.3816A}. 

In a search for metal-rich \gse\ stars, \citet{2021SCPMA..6439562Z} compare the maximum height above or below the Galactic plane ($z_{\mathrm{max}}$) to the mean Galactic distance ($R_{\mathrm{m}} = R_{\mathrm{apo}} + R_{\mathrm{peri}} / 2$, where $R_{\mathrm{apo}}$ is the apocenter distance and $R_{\mathrm{peri}}$ is the pericenter distance) for a sample of stars that includes the star G21-22, after which the putative G21-22 moving group was named. The features of the resulting \citet{2021SCPMA..6439562Z} plot (their figure 7) are similar to those found by \citet{2020MNRAS.492.3816A}, who compare $z_{\mathrm{max}}$ to $R_{\mathrm{apo}}$ (their figure B1). 

\citet{2020MNRAS.492.3816A} use newly derived stellar halo and thick-disk fractions, and new stellar kinematics with the Galaxia population synthesis model \citep{2011ApJ...730....3S}, to conclude that this $z_{\mathrm{max}}$-$R_{\mathrm{apo}}$ relation is a natural consequence of resonant effects. However, \citet{2018ApJ...863..113H} examine the blue and red sequences of the \gaia\ HR diagram \citet{2018A&A...616A..10G} and find a similar relation for $z_{\mathrm{max}}$-$R_{\mathrm{R2D,max}}$ (the apocenter distance projected on the Galactic plane) and, based on this and other dynamical and orbital properties, conclude the blue sequence stars are possible relics of a significant accretion event experienced by the Galaxy.
    
We attempted to determine the origins of the G21-22 stars by comparing their stellar properties to other stars in the Milky Way halo in kinematic and chemical space. While it is generally accepted that at least part of the halo has been populated by accretion events \citep[e.g.,][]{2020ARA&A..58..205H}, some recent studies suggest that it was built entirely by accretion \citep{2017A&A...598A..58H,2020ApJ...901...48N}. This raises the possibility that the G21-22 stars can be associated to one or more of these structures, each of which occupies a characteristic region in integrals of motion, particularly $E$-$L_{\mathrm{z}}$ \citep{2020ApJ...901...48N}.

    \begin{figure}
    \begin{center}
    \scalebox{0.35}{\includegraphics{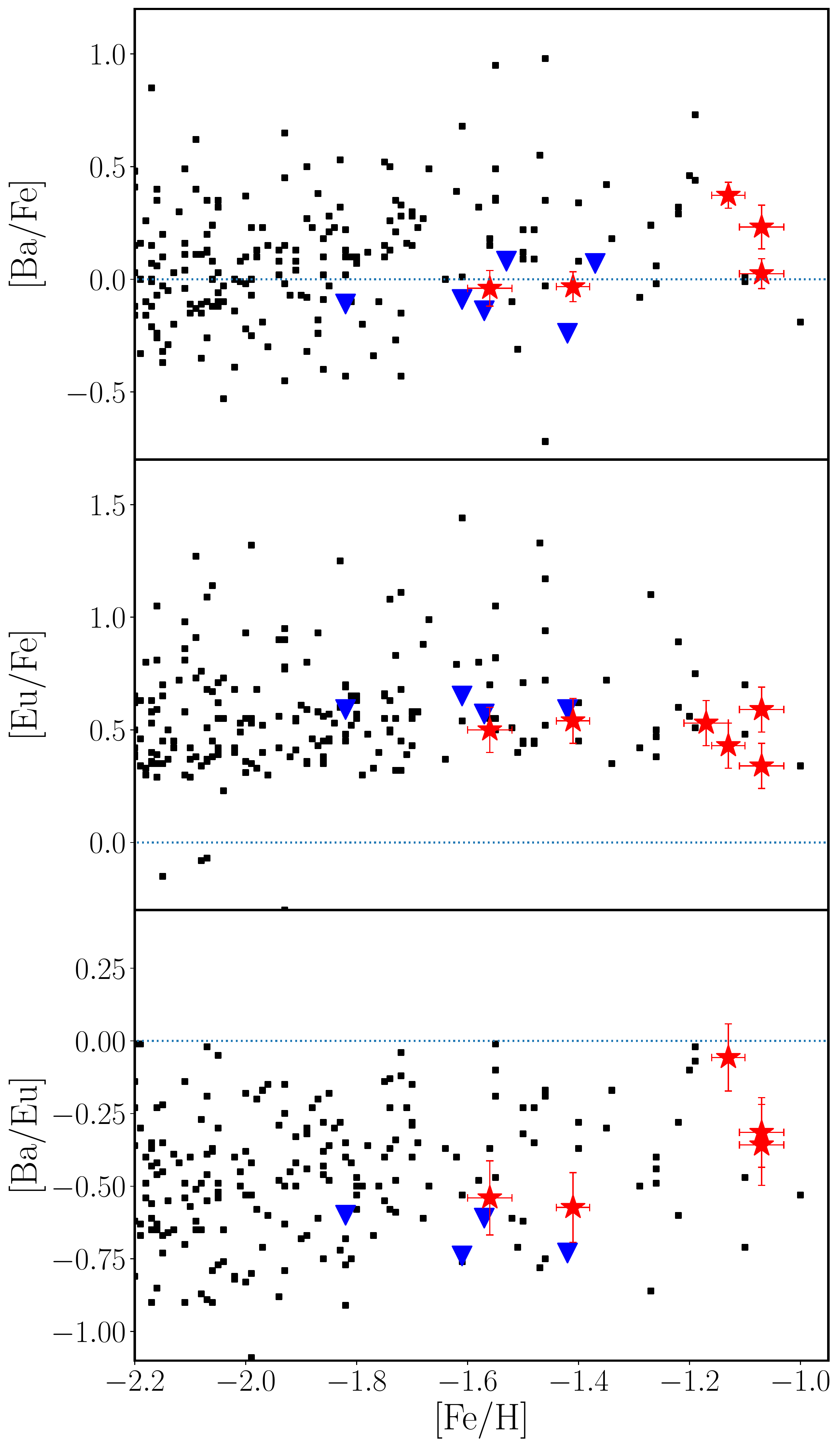}}
    \caption{Abundances of Ba, Eu, and Ba relative to Eu as a function of metallicity. Symbols are the same as given in Figure~\ref{fig:alpha} with the following sources: \gse-- \citet{2020MNRAS.497.1236M} and \citet{2021ApJ...908L...8A}; \textit{Halo}-- \citet{2021ApJ...908...79G}; \textit{Sagittarius}-- \citet{2018ApJ...855...83H}.  \label{fig:baeu}}
    \end{center}
    \end{figure}

\citet{2020ApJ...901...48N} derived their integrals of motion using the same Galactic model as we did (see Section~\ref{sec:gaia}). Using these authors' results as a guide, we can eliminate Sagittarius, Sequoia, Helmi stream, and Thamnos as progenitors of the six G21-22 stars based on their positions in the $E$-$L_{\mathrm{z}}$ and $V_{\phi}$-$V_{\mathrm{r}}$ planes. There is no overlap between these structures and the stars. 

The kinematic properties of the stars are consistent with the lower density regions of the high-$\alpha$ disk and in situ halo component of \citet{2020ApJ...901...48N}, but they do not meet the metalicity cutoff set by these authors to distinguish it from the metal-weak thick disk (MWTD), which is taken to be the metal-poor extension of the high-$\alpha$ disk. Conversely, two of the G21-22 stars fall in the [$\alpha$/Fe]\footnote{We calculate [$\alpha$/Fe] as the average abundance of the $\alpha$-elements Mg, Si, Ca, and Ti.}-[Fe/H] selection box set for the MWTD, but as we show in Figure~\ref{fig:L_z_E}, the stars are clearly part of the halo and are not consistent with the MWTD in any of the kinematic spaces. 

This leaves \gse, which has been tied to greater than 50\% of the Galactic halo \citep{2019MNRAS.482.3426M}, as the possible progenitor. Indeed, the six G21-22 stars are consistent with \gse\ in four key diagnostics presented by \citet{2020ApJ...901...48N}: they have eccentricities $e \geq 0.77$ (see Table~\ref{tab:2}), which is larger than the cutoff value of $e = 0.7$, and they overlap with \gse\ stars in the $E$-$L_{\mathrm{z}}$, $V_{\phi}$-$V_{\mathrm{r}}$, and [$\alpha$/Fe]-[Fe/H] spaces. Moreover, these stars' metallicities ([Fe/H]) are within about $1\sigma$ of the mean metallicty of the structure, depending on the value considered \citep[e.g.,][]{2019MNRAS.487L..47V,2019ApJ...874L..35M,2020ApJ...901...48N}.
    
To investigate the six G21-22 stars in chemical space, their abundances are plotted with those of general halo populations and \ocen\ in Figures~\ref{fig:alpha}, \ref{fig:fePeak}, \ref{fig:oddZ}, and \ref{fig:baeu}. The general halo population includes stars observed by APOGEE and deemed by \citet{2018ApJ...852...49H} to have likely been accreted by the Milky Way early in its history (labeled the low-Mg population, LMg) and to have formed in situ, possibly as part of the thick disk (labeled the high-Mg population, HMg). The remaining halo stars are found in \citet{2004AJ....128.1177V}, \citet{2006MNRAS.367.1329R}, \citet{2010A&A...511L..10N}, \citet{2012ApJ...753...64I}, \citet{2013ApJ...771...67I}, and \citet{2014AJ....147..136R}. Abundances of \ocen\ stars are from \citet{2010ApJ...722.1373J}. 
    
The abundances of \ocen, LMg, and HMg occupy distinct regions of the abundance plots, although there is some overlap between LMg and both \ocen\ (at low-[Fe/H]) and HMg (at high-[Fe/H]) for some elements. The G21-22 stars follow the general halo trend, which spans across the \ocen, LMg, and HMg stars, and cannot be definitively assigned to any of the three. The abundances of the two more metal-poor stars-- G02-38 and G119-64-- are consistent with those of \ocen, raising the possibility that they are part of the cluster's tidal debris, as suggested by \citet{2012RMxAA..48..109S}. However, $L_{\mathrm{z}}$ of these stars are inconsistent with the recently discovered Fimbulthul stream, which has been identified as the tidal stream of \ocen\ \citep{2019ApJ...872..152I,2019NatAs...3..667I}, arguing against an association with the cluster. 

The four more metal-rich stars, G82-42, G108-48, G161-73, and HIP~36878, have abundances generally consistent with those of both the LMg and HMg groups. However, given that the kinematics of all six stars are consistent with \gse\ and much of the general halo is composed of stars from \gse, possibly even the LMg stars, we suggest it as the more likely progenitor of the G21-22 stars. If this is indeed the case, the abundances presented here will add to the nascent list of abundances derived for \gse\ stars.

\section{Conclusions} \label{sec:conclusion} 
We used \gaia\ astrometry and detailed stellar abundances derived from high-resolution VLT/UVES spectroscopy to investigate the authenticity of three candidate halo moving groups-- G03-37, G18-39, and G21-22-- identified by \citet{2012RMxAA..48..109S}. The stars from each group evince significant scatter in integrals of motion (energy, angular momentum) and velocities, contrary to the tight clustering seen for members of moving groups. Moreover, stellar orbits of the stars integrated backward in a Galactic potential over 200~Myr are found to deviate from kinematically coherent orbits, suggesting the stars did not have common space motions in the past. These kinematic diagnostics are counter to what is expected for bona fide moving group stars.

Choosing the putative moving group showing the least scatter in the integrals of motion, we conducted a detailed abundance analysis of the six stars in our G21-22 sample to further investigate the nature of this group. We derived the abundances of 14 elements, including $\alpha$-, Fe-peak, odd-$Z$, and neutron-capture elements, and find that the stars are not chemically homogeneous. Therefore, using the definition of a moving group as a stellar association with a shared formation history, shared space motions, and shared compositions, we concluded that G03-37, G18-39, and G21-22 are not genuine moving groups based on their assigned stars' discordance in the integrals of motion, integrated orbits, and in the case of G21-22, compositions. 

By comparing the integrals of motion of the six stars of G21-22 to those of known and newly discovered structures in the halo presented in \citet{2020ApJ...901...48N}, we tentatively associated the G21-22 stars with the \gse\ accretion event. They overlap with \gse\ in five key kinematic and chemical diagnostics, while being inconsistent with other major halo structures. If their association is confirmed, the detailed abundances presented here will be valuable in the continued characterization of \gse\ and the evolution of the Galactic halo.

While integrals of motion are well established diagnostics for identifying large stellar structures accreted by the Galaxy \citep[e.g.,][]{2000MNRAS.319..657H,2019MNRAS.490L..32S}, uncertainties in astrometric parameters can lead to fallacious kinematic quantities that result in uncertain conclusions for smaller associations such as moving groups. Detailed stellar abundances can be used to chemically tag groups of stars, but for stars with disparate positions in the Galaxy-- like moving group stars-- similar compositions alone may be insufficient to definitely conclude that they share a common formation history. As we have demonstrated, however, the combination of kinematic and chemical information is much more effective at identifying truly conatal populations than either alone.

\begin{acknowledgments}
The authors thank Chris Sneden for helpful discussions on MOOG, Eu abundances, and the development of the model atmosphere interpolator embedded in \spae. SCS, VRC, and MM were supported by an Undergraduate Research and Inquiry (URI) Grant from the University of Tampa. SCS was also supported by a Research Innovation and Scholarly Excellence (RISE) Grant from the University of Tampa. J.J.A.\ acknowledges support from CIERA and Northwestern University through a Postdoctoral Fellowship. J.C. acknowledges support from the Agencia Nacional de Investigación y Desarrollo (ANID) via Proyecto Fondecyt Regular 1191366; and from "Centro de Astronomía y Tecnologías Afines" project BASAL AFB-170002.
\end{acknowledgments}

\software{\spae\ \citep{spae}, {\tt astropy} \citep{astropy:2013,astropy:2018}, {\tt gala} \citep{gala_zenodo}, {\tt NumPy} \citep{harris2020array}, {\tt SciPy} \citep{2020SciPy-NMeth}, {\tt matplotlib} \citep{matplotlib}, {\tt corner} \citep{corner}, {\tt emcee} \citep{2013PASP..125..306F}}

\newpage
\appendix
\restartappendixnumbering
\section{\gaia\ EDR3 Astrometry and Derived Kinematic Properties of G03-37, G18-39, and G21-22 \label{append:gaia}}
Table \ref{tab:1} presents the complete six-dimensional phase space information from the early third \gaia\ data release \citep[EDR3;][]{2021A&A...649A...1G, 2021A&A...649A...2L} for each star identified by \citet{2012RMxAA..48..109S} for the three candidate moving groups G03-37, G18-39, and G21-22. The stars were crossed matched in \gaia\ EDR3 using identifications and positional data from \citet{2012RMxAA..48..109S} and \citet{schuster}.

\startlongtable
\begin{deluxetable*}{lDDDDDD}
\tablecolumns{13}
\tablewidth{0pt}
\tablecaption{Observed Gaia EDR3 Characteristics of Halo Moving Group Stars \label{tab:1}}
\tablehead{
	\colhead{Star}&
	\twocolhead{$\alpha$}&
	\twocolhead{$\delta$}&
	\twocolhead{$\mu_{\alpha}$}&
	\twocolhead{$\mu_{\delta}$}&
	\twocolhead{$\varpi$}&
	\twocolhead{RV}\\
	\colhead{}&
	\twocolhead{}&
	\twocolhead{}&
	\twocolhead{(mas yr$^{-1}$)}&
    \twocolhead{(mas yr$^{-1}$)}&
    \twocolhead{(mas)}&
    \twocolhead{(km s$^{-1}$)}
	}
\decimals
\startdata
\cutinhead{G03-37}
LP659-016 & 06:07:22.84 & $-$07:10:31.0 & $140.85\pm0.05$  & $-120.93\pm0.05$ & $4.18\pm0.05$  & $179.6\pm2.1$ \\
LP685-044 & 16:41:57.4  & $-$07:52:36.8 & $-236.17\pm0.02$ & $-166.37\pm0.02$ & $4.28\pm0.02$  & \nodata \\
G03-37    & 02:03:02.23 & $+$08:51:56.4 & $143.31\pm0.02$  & $-98.64\pm0.02$  & $2.45\pm0.02$  & $-83.0\pm1.6$ \\
G05-36    & 03:27:00.07 & $+$23:46:30.4 & $263.00\pm0.02$  & $-344.70\pm0.01$ & $5.76\pm0.01$  & $-9.1\pm0.3$ \\
G51-07    & 08:20:06.83 & $+$34:42:10.6 & $14.54\pm0.01$   & $-377.11\pm0.01$ & $8.36\pm0.02$  & $80.9\pm3.9$ \\
G53-41    & 10:27:24.04 & $+$01:23:55.4 & $-186.34\pm0.02$ & $-294.81\pm0.02$ & $6.06\pm0.02$  & $88.8\pm0.5$ \\
G142-44   & 19:38:52.9  & $+$16:25:31.1 & $-197.36\pm0.03$ & $-185.98\pm0.03$ & $9.05\pm0.03$  & \nodata \\
G170-56   & 17:38:15.39 & $+$18:33:22.2 & $-187.82\pm0.01$ & $-203.90\pm0.01$ & $7.68\pm0.01$  & $-239.4\pm0.5$ \\
G172-61   & 01:34:23.28 & $+$48:44:26.5 & $363.98\pm0.17$  & $-57.95\pm0.16$  & $9.44\pm0.17$  & $-203.4\pm1.9$ \\
G272-122  & 01:54:32.33 & $-$18:27:13.0 & $169.11\pm0.01$  & $-147.24\pm0.01$ & $3.61\pm0.02$  & $-15.3\pm2.0$ \\
HIP~22068  & 04:44:49.13 & $-$32:52:40.0 & $352.44\pm0.01$  & $-190.26\pm0.02$ & $7.20\pm0.01$  & $150.2\pm0.3$ \\
HIP~59490  & 12:12:01.12 & $+$13:15:33.0 & $-218.30\pm0.03$ & $-439.48\pm0.02$ & $7.45\pm0.03$  & $99.5\pm0.3$ \\
HIP~62108  & 12:43:42.89 & $-$44:40:31.8 & $-220.73\pm0.02$ & $-17.45\pm0.01$  & $6.80\pm0.02$  & $222.6\pm0.8$ \\
HIP~69232  & 14:10:26.70 & $-$13:56:11.6 & $-339.94\pm0.04$ & $-459.56\pm0.04$ & $7.11\pm0.04$  & $120.2\pm1.7$ \\
LTT5864   & 14:46:27.05 & $-$35:21:31.5 & $-465.91\pm0.02$ & $-246.62\pm0.02$ & $11.19\pm0.02$ & $-139.9\pm1.2$ \\
\cutinhead{G18-39}
BD-035166 & 21:17:18.91 & $-$02:44:58.7 & $-101.40\pm0.02$ & $-243.28\pm0.01$ & $3.91\pm0.02$  & $-126.9\pm0.4$ \\
CD-249840 & 11:36:02.67 & $-$24:52:25.0 & $-17.91\pm0.02$  & $-211.86\pm0.02$ & $3.63\pm0.02$  & $164.7\pm0.7$ \\
CD-482445 & 06:41:26.76 & $-$48:13:11.4 & $65.05\pm0.02$   & $217.54\pm0.02$  & $5.37\pm0.01$  & $320.8\pm0.5$ \\
LP490-061 & 10:43:59.20 & $+$12:48:03.8 & $-147.43\pm0.04$ & $-97.59\pm0.03$  & $2.38\pm0.04$  & $216.3\pm3.5$ \\
LP636-003 & 20:45:01.20 & $-$01:40:58.4 & $-34.78\pm0.02$  & $-179.62\pm0.02$ & $2.57\pm0.02$  & \nodata \\
LP709-053 & 02:12:02.41 & $-$14:00:29.7 & $266.30\pm0.03$  & $-63.26\pm0.02$  & $3.65\pm0.03$  & $-31.4\pm1.3$ \\
LP877-025 & 22:53:29.9  & $-$23:54:53.6 & $-30.32\pm0.03$  & $-396.54\pm0.03$ & $5.53\pm0.04$  & $81.0\pm1.0$ \\
G10-03    & 11:10:02.8  & $-$02:47:33.0 & $144.92\pm0.02$  & $-472.29\pm0.01$ & $9.91\pm0.02$  & $203.1\pm0.6$ \\
G15-24    & 15:30:41.34 & $+$08:23:38.6 & $-391.30\pm0.02$ & $-116.91\pm0.01$ & $6.56\pm0.02$  & $-88.5\pm0.5$ \\
G18-39    & 22:18:36.81 & $+$08:26:43.3 & $283.28\pm0.02$  & $-102.14\pm0.02$ & $6.38\pm0.02$  & $-233.9\pm0.6$ \\
G31-26    & 00:08:20.03 & $-$05:14:57.0 & $352.95\pm0.03$  & $-131.29\pm0.02$ & $5.16\pm0.03$  & $-216.8\pm1.4$ \\
G34-45    & 01:42:48.85 & $+$22:36:57.0 & $74.07\pm0.02$   & $-301.57\pm0.01$ & $7.07\pm0.02$  & $-272.1\pm6.9$ \\
G36-47    & 02:57:20.15 & $+$26:16:49.7 & $259.58\pm0.02$  & $-226.55\pm0.02$ & $4.66\pm0.03$  & $89.5\pm0.4$ \\
G66-30    & 14:50:07.49 & $+$00:50:25.5 & $-291.58\pm0.15$ & $-101.42\pm0.13$ & $6.65\pm0.14$  & $-114.5\pm5.1$ \\
G70-33    & 01:03:54.63 & $-$03:51:15.3 & $339.62\pm0.02$  & $-64.68\pm0.02$  & $4.14\pm0.02$  & $-82.5\pm3.0$ \\
G86-39    & 05:23:12.04 & $+$33:10:51.3 & $391.73\pm0.02$  & $-655.36\pm0.02$ & $12.66\pm0.02$ & $214.7\pm0.6$ \\
G87-13    & 06:54:56.36 & $+$35:30:54.8 & $52.69\pm0.02$   & $-237.88\pm0.02$ & $4.06\pm0.02$  & $206.6\pm2.5$ \\
G89-14    & 07:22:31.61 & $+$08:49:08.7 & $151.20\pm0.03$  & $-269.47\pm0.02$ & $4.40\pm0.03$  & $-38.5\pm1.7$ \\
G98-53    & 06:13:49.85 & $+$33:24:56.9 & $23.21\pm0.02$   & $-325.84\pm0.01$ & $5.46\pm0.02$  & $144.8\pm1.0$ \\
G120-15   & 11:06:20.13 & $+$31:12:47.5 & $-194.82\pm0.02$ & $-129.13\pm0.02$ & $2.98\pm0.02$  & $130.3\pm0.3$ \\
G157-85   & 23:36:31.51 & $-$08:25:54.9 & $260.95\pm0.06$  & $-146.97\pm0.05$ & $3.95\pm0.06$  & $-69.1\pm2.0$ \\
G176-53   & 11:46:33.68 & $+$50:52:45.9 & $-868.72\pm0.01$ & $-544.10\pm0.01$ & $14.16\pm0.02$ & $64.9\pm0.4$ \\
G187-30   & 21:11:17.69 & $+$33:31:29.8 & $502.72\pm0.01$  & $159.62\pm0.02$  & $10.18\pm0.02$ & $-340.2\pm0.7$ \\
HD~3567    & 00:38:31.96 & $-$08:18:42.1 & $20.63\pm0.03$   & $-546.54\pm0.03$ & $8.52\pm0.02$  & $-46.8\pm0.3$ \\
HD~101063  & 11:37:40.02 & $-$28:51:05.1 & $-313.63\pm0.01$ & $-15.67\pm0.01$  & $4.11\pm0.02$  & $183.2\pm0.4$ \\
\cutinhead{G21-22}
LP808-022 & 17:50:36.52 & $-$16:59:06.1 & $-51.84\pm0.02$  & $-213.58\pm0.02$ & $3.02\pm0.02$ & \nodata \\
G02-38    & 01:26:55.15 & $+$12:00:20.0 & $-12.21\pm0.03$  & $-358.99\pm0.02$ & $5.49\pm0.02$ & $-171.9\pm0.7$ \\
G21-22    & 18:39:09.54 & $+$00:07:07.1 & $-163.73\pm0.02$ & $-445.51\pm0.01$ & $6.69\pm0.02$ & $58.9\pm0.6$ \\
G82-42    & 04:42:00.01 & $-$04:16:52.0 & $1.92\pm0.01$    & $-344.68\pm0.01$ & $4.78\pm0.01$ & $-7.8\pm0.9$ \\
G108-48   & 07:01:36.89 & $+$06:24:17.9 & $11.39\pm0.02$   & $-676.01\pm0.01$ & $8.12\pm0.02$ & $-86.9\pm0.4$ \\
G116-45   & 09:44:38.16 & $+$38:36:34.5 & $212.25\pm0.02$  & $-239.66\pm0.02$ & $4.39\pm0.02$ & $-32.9\pm0.9$ \\
G119-64   & 11:12:48.09 & $+$35:43:35.7 & $71.00\pm0.02$   & $-510.84\pm0.02$ & $7.28\pm0.02$ & $-195.0\pm0.4$ \\
G139-16   & 17:09:47.21 & $+$08:04:19.9 & $-160.03\pm0.01$ & $-362.10\pm0.01$ & $5.08\pm0.01$ & $39.7\pm1.2$ \\
G146-56   & 10:39:50.60 & $+$42:09:01.7 & $87.08\pm0.02$   & $-208.15\pm0.01$ & $3.56\pm0.01$ & \nodata \\
G154-25   & 17:50:36.52 & $-$16:59:06.1 & $-51.84\pm0.02$  & $-213.58\pm0.02$ & $3.02\pm0.02$ & \nodata \\
G161-73   & 09:45:37.99 & $-$04:40:32.5 & $152.15\pm0.02$  & $-250.73\pm0.02$ & $4.37\pm0.02$ & $120.4\pm0.7$ \\
HIP~36878  & 07:34:53.41 & $-$10:23:17.7 & $424.66\pm0.01$  & $-481.99\pm0.01$ & $9.04\pm0.02$ & $80.3\pm0.8$ \\
\enddata

\end{deluxetable*}

Table \ref{tab:2} presents veloicities and integrals of motion for each star identified by \citet{2012RMxAA..48..109S} for the three candidate moving groups G03-37, G18-39, and G21-22. The quantities have been derived using the Galactic modeling framework of {\tt Astropy v4.0} \citep{astropy:2013,astropy:2018} and the Galactic modeling code, {\tt gala} \citep{2017JOSS....2..388P}, with its built-in Galactic potential {\tt MilkyWayPotential} (see Table~\ref{tab:2}). This model combines the disk model from \citet{2015ApJS..216...29B} with a spherical NFW profile \citep{1997ApJ...490..493N}.

\startlongtable
\begin{deluxetable*}{lDDDDDDDDD}
\tabletypesize{\footnotesize}
\tablecolumns{19}
\tablewidth{0pt}
\tablecaption{Derived Characteristics of Halo Moving Group Stars \label{tab:2}}
\tablehead{
	\colhead{Star}&
	\twocolhead{$V_{\rho}$}&
	\twocolhead{$V_{z}$}&
	\twocolhead{$V_{\phi}$}&
	\twocolhead{$U$}&
	\twocolhead{$V$}&
	\twocolhead{$W$}&
	\twocolhead{E}&
	\twocolhead{$L_z$}&
	\twocolhead{$e$}\\
	\colhead{}&
	\twocolhead{(km~s$^{-1}$)}&
    \twocolhead{(km~s$^{-1}$)}&
	\twocolhead{(km~s$^{-1}$)}&
    \twocolhead{(km~s$^{-1}$)}&
	\twocolhead{(km~s$^{-1}$)}&
    \twocolhead{(km~s$^{-1}$)}&
    \twocolhead{($10^5$~km$^2$~s$^{-2}$)}&
    \twocolhead{($10^3$~kpc~km~s$^{-1}$)}&
    \twocolhead{}
	}
\decimals
\startdata
\cutinhead{G03-37}
LP659-016 & 37.8 & 47.7 & 22.8 & -40.4 & -263.7 & 47.0 & -1.53 & 0.189 & 0.90 \\
LP685-044 & \nodata & \nodata & \nodata & \nodata & \nodata & \nodata & \nodata & \nodata & \nodata \\
G03-37 & 62.8 & 36.5 & 92.7 & -67.1 & -332.1 & 35.7 & -1.48 & 0.774 & 0.62 \\
G05-36 & 43.4 & -64.6 & 99.6 & -46.7 & -339.7 & -65.4 & -1.47 & 0.824 & 0.57 \\
G51-07 & 48.4 & 18.8 & -25.6 & -51.4 & -214.8 & 18.0 & -1.54 & -0.211 & 0.90 \\
G53-41 & 25.0 & -104.7 & 15.6 & -27.4 & -256.2 & -105.4 & -1.50 & 0.127 & 0.79 \\
G142-44 & \nodata & \nodata & \nodata & \nodata & \nodata & \nodata & \nodata & \nodata & \nodata \\
G170-56 & 59.0 & -35.4 & 37.0 & -62.1 & -276.8 & -36.2 & -1.54 & 0.297 & 0.84 \\
G172-61 & 0.8 & 57.8 & 24.4 & -4.1 & -264.8 & 57.2 & -1.54 & 0.200 & 0.88 \\
G272-122 & 29.8 & 45.1 & 44.0 & -32.7 & -284.4 & 44.3 & -1.53 & 0.361 & 0.81 \\
HIP~22068 & -0.5 & 71.2 & 50.9 & -2.0 & -291.2 & 70.6 & -1.52 & 0.416 & 0.76 \\
HIP~59490 & -28.6 & 14.0 & 81.8 & 26.1 & -322.0 & 13.5 & -1.52 & 0.665 & 0.66 \\
HIP~62108 & 4.8 & 61.1 & 19.3 & -7.5 & -259.7 & 60.4 & -1.55 & 0.155 & 0.90 \\
HIP~69232 & -31.8 & -37.0 & 151.3 & 30.0 & -391.4 & -37.6 & -1.44 & 1.215 & 0.37 \\
LTT5864 & 214.6 & -45.3 & -126.1 & -218.1 & -115.5 & -46.5 & -1.25 & -1.015 & 0.71 \\
\cutinhead{G18-39}
BD-035166 & -165.5 & 38.3 & 58.6 & 161.3 & -302.2 & 38.1 & -1.41 & 0.468 & 0.82 \\
CD-249840 & -145.3 & -117.5 & 24.5 & 143.3 & -260.9 & -117.8 & -1.39 & 0.198 & 0.81 \\
CD-482445 & 225.4 & -5.1 & 44.1 & -227.3 & -289.1 & -6.3 & -1.30 & 0.360 & 0.90 \\
LP490-061 & 227.7 & -2.5 & 87.4 & -228.6 & -332.7 & -3.7 & -1.25 & 0.722 & 0.81 \\
LP636-003 & \nodata & \nodata & \nodata & \nodata & \nodata & \nodata & \nodata & \nodata & \nodata \\
LP709-053 & 168.5 & 123.4 & 38.8 & -171.7 & -279.3 & 122.4 & -1.33 & 0.320 & 0.75 \\
LP877-025 & -188.7 & -88.6 & 41.0 & 185.9 & -282.4 & -88.7 & -1.34 & 0.330 & 0.84 \\
G10-03 & -160.8 & 80.8 & 19.8 & 157.9 & -259.0 & 80.6 & -1.40 & 0.161 & 0.91 \\
G15-24 & 132.4 & 73.8 & 18.3 & -135.5 & -258.2 & 72.8 & -1.45 & 0.147 & 0.87 \\
G18-39 & 176.4 & -3.5 & 19.0 & -179.6 & -256.8 & -4.6 & -1.41 & 0.154 & 0.95 \\
G31-26 & 200.4 & 107.0 & 89.8 & -204.4 & -328.2 & 105.8 & -1.26 & 0.730 & 0.72 \\
G34-45 & -190.7 & 34.8 & 44.6 & 187.3 & -286.6 & 34.7 & -1.36 & 0.366 & 0.88 \\
G36-47 & 179.5 & -83.6 & 48.6 & -182.6 & -287.2 & -84.7 & -1.34 & 0.403 & 0.82 \\
G66-30 & 149.0 & -21.7 & -60.1 & -151.9 & -180.4 & -22.7 & -1.44 & -0.483 & 0.80 \\
G70-33 & 248.2 & 73.7 & 57.4 & -251.8 & -295.6 & 72.4 & -1.21 & 0.470 & 0.86 \\
G86-39 & 230.7 & -15.6 & 15.3 & -233.5 & -255.4 & -16.9 & -1.29 & 0.126 & 0.97 \\
G87-13 & 195.0 & 19.2 & 37.5 & -197.9 & -277.9 & 18.1 & -1.35 & 0.313 & 0.90 \\
G89-14 & -207.5 & 18.2 & 29.2 & 204.9 & -266.9 & 18.1 & -1.33 & 0.243 & 0.93 \\
G98-53 & 149.8 & -85.6 & 11.7 & -152.4 & -252.0 & -86.7 & -1.40 & 0.097 & 0.89 \\
G120-15 & 234.9 & 10.8 & 60.0 & -237.5 & -301.5 & 9.6 & -1.26 & 0.495 & 0.87 \\
G157-85 & 184.1 & -76.6 & 52.7 & -187.4 & -290.6 & -77.8 & -1.35 & 0.427 & 0.81 \\
G176-53 & 209.5 & 62.3 & 18.5 & -212.6 & -258.4 & 61.2 & -1.32 & 0.151 & 0.91 \\
G187-30 & 263.0 & -54.1 & 66.9 & -266.5 & -304.1 & -55.5 & -1.18 & 0.542 & 0.87 \\
HD~3567 & -163.5 & -47.1 & 16.7 & 160.6 & -257.8 & -47.4 & -1.42 & 0.136 & 0.93 \\
HD~101063 & 259.3 & -12.9 & 49.9 & -260.9 & -296.7 & -14.2 & -1.22 & 0.403 & 0.91 \\
\cutinhead{G21-22}
LP808-022 & \nodata & \nodata & \nodata & \nodata & \nodata & \nodata & \nodata & \nodata & \nodata \\
G02-38 & -224.2 & -57.7 & 29.6 & 221.2 & -272.2 & -57.8 & -1.29 & 0.243 & 0.90 \\
G21-22 & -239.7 & -30.7 & 4.2 & 236.8 & -246.9 & -30.7 & -1.28 & 0.034 & 0.99 \\
G82-42 & -204.6 & -130.6 & 2.8 & 202.1 & -241.6 & -130.7 & -1.26 & 0.023 & 0.87 \\
G108-48 & -243.6 & -171.8 & 35.0 & 241.4 & -273.6 & -171.8 & -1.11 & 0.288 & 0.77 \\
G116-45 & -234.6 & 137.1 & -11.0 & 231.3 & -229.0 & 137.0 & -1.18 & -0.091 & 0.85 \\
G119-64 & -233.9 & -121.6 & 46.6 & 231.4 & -286.8 & -121.7 & -1.20 & 0.381 & 0.83 \\
G139-16 & -227.4 & 10.4 & 56.0 & 223.8 & -298.7 & 10.3 & -1.30 & 0.446 & 0.87 \\
G146-56 & \nodata & \nodata & \nodata & \nodata & \nodata & \nodata & \nodata & \nodata & \nodata \\
G154-25 & \nodata & \nodata & \nodata & \nodata & \nodata & \nodata & \nodata & \nodata & \nodata \\
G161-73 & -230.8 & 45.6 & 17.1 & 228.1 & -252.8 & 45.5 & -1.28 & 0.140 & 0.95 \\
HIP~36878 & -203.5 & 87.6 & 34.0 & 200.7 & -272.4 & 87.4 & -1.31 & 0.279 & 0.87 \\
\enddata

\end{deluxetable*}

\bibliographystyle{aasjournal}
\bibliography{g2122}

\begin{thebibliography}{}
\expandafter\ifx\csname natexlab\endcsname\relax\def\natexlab#1{#1}\fi
\providecommand{\url}[1]{\href{#1}{#1}}
\providecommand{\dodoi}[1]{doi:~\href{http://doi.org/#1}{\nolinkurl{#1}}}
\providecommand{\doeprint}[1]{\href{http://ascl.net/#1}{\nolinkurl{http://ascl.net/#1}}}
\providecommand{\doarXiv}[1]{\href{https://arxiv.org/abs/#1}{\nolinkurl{https://arxiv.org/abs/#1}}}

\bibitem[{{Aguado} {et~al.}(2021){Aguado}, {Belokurov}, {Myeong}, {Evans},
  {Kobayashi}, {Sbordone}, {Chanam{\'e}}, {Navarrete}, \&
  {Koposov}}]{2021ApJ...908L...8A}
{Aguado}, D.~S., {Belokurov}, V., {Myeong}, G.~C., {et~al.} 2021, \apjl, 908,
  L8, \dodoi{10.3847/2041-8213/abdbb8}

\bibitem[{{Amarante} {et~al.}(2020){Amarante}, {Smith}, \&
  {Boeche}}]{2020MNRAS.492.3816A}
{Amarante}, J. A.~S., {Smith}, M.~C., \& {Boeche}, C. 2020, \mnras, 492, 3816,
  \dodoi{10.1093/mnras/staa077}

\bibitem[{{Asplund} {et~al.}(2021){Asplund}, {Amarsi}, \&
  {Grevesse}}]{2021arXiv210501661A}
{Asplund}, M., {Amarsi}, A.~M., \& {Grevesse}, N. 2021, arXiv e-prints,
  arXiv:2105.01661.
\newblock \doarXiv{2105.01661}

\bibitem[{{Asplund} {et~al.}(2009){Asplund}, {Grevesse}, {Sauval}, \&
  {Scott}}]{2009ARA&A..47..481A}
{Asplund}, M., {Grevesse}, N., {Sauval}, A.~J., \& {Scott}, P. 2009, \araa, 47,
  481, \dodoi{10.1146/annurev.astro.46.060407.145222}

\bibitem[{{Astropy Collaboration} {et~al.}(2013){Astropy Collaboration},
  {Robitaille}, {Tollerud}, {Greenfield}, {Droettboom}, {Bray}, {Aldcroft},
  {Davis}, {Ginsburg}, {Price-Whelan}, {Kerzendorf}, {Conley}, {Crighton},
  {Barbary}, {Muna}, {Ferguson}, {Grollier}, {Parikh}, {Nair}, {Unther},
  {Deil}, {Woillez}, {Conseil}, {Kramer}, {Turner}, {Singer}, {Fox}, {Weaver},
  {Zabalza}, {Edwards}, {Azalee Bostroem}, {Burke}, {Casey}, {Crawford},
  {Dencheva}, {Ely}, {Jenness}, {Labrie}, {Lim}, {Pierfederici}, {Pontzen},
  {Ptak}, {Refsdal}, {Servillat}, \& {Streicher}}]{astropy:2013}
{Astropy Collaboration}, {Robitaille}, T.~P., {Tollerud}, E.~J., {et~al.} 2013,
  \aap, 558, A33, \dodoi{10.1051/0004-6361/201322068}

\bibitem[{{Astropy Collaboration} {et~al.}(2018){Astropy Collaboration},
  {Price-Whelan}, {Sip{\H{o}}cz}, {G{\"u}nther}, {Lim}, {Crawford}, {Conseil},
  {Shupe}, {Craig}, {Dencheva}, {Ginsburg}, {Vand erPlas}, {Bradley},
  {P{\'e}rez-Su{\'a}rez}, {de Val-Borro}, {Aldcroft}, {Cruz}, {Robitaille},
  {Tollerud}, {Ardelean}, {Babej}, {Bach}, {Bachetti}, {Bakanov}, {Bamford},
  {Barentsen}, {Barmby}, {Baumbach}, {Berry}, {Biscani}, {Boquien}, {Bostroem},
  {Bouma}, {Brammer}, {Bray}, {Breytenbach}, {Buddelmeijer}, {Burke},
  {Calderone}, {Cano Rodr{\'\i}guez}, {Cara}, {Cardoso}, {Cheedella}, {Copin},
  {Corrales}, {Crichton}, {D'Avella}, {Deil}, {Depagne}, {Dietrich}, {Donath},
  {Droettboom}, {Earl}, {Erben}, {Fabbro}, {Ferreira}, {Finethy}, {Fox},
  {Garrison}, {Gibbons}, {Goldstein}, {Gommers}, {Greco}, {Greenfield},
  {Groener}, {Grollier}, {Hagen}, {Hirst}, {Homeier}, {Horton}, {Hosseinzadeh},
  {Hu}, {Hunkeler}, {Ivezi{\'c}}, {Jain}, {Jenness}, {Kanarek}, {Kendrew},
  {Kern}, {Kerzendorf}, {Khvalko}, {King}, {Kirkby}, {Kulkarni}, {Kumar},
  {Lee}, {Lenz}, {Littlefair}, {Ma}, {Macleod}, {Mastropietro}, {McCully},
  {Montagnac}, {Morris}, {Mueller}, {Mumford}, {Muna}, {Murphy}, {Nelson},
  {Nguyen}, {Ninan}, {N{\"o}the}, {Ogaz}, {Oh}, {Parejko}, {Parley}, {Pascual},
  {Patil}, {Patil}, {Plunkett}, {Prochaska}, {Rastogi}, {Reddy Janga},
  {Sabater}, {Sakurikar}, {Seifert}, {Sherbert}, {Sherwood-Taylor}, {Shih},
  {Sick}, {Silbiger}, {Singanamalla}, {Singer}, {Sladen}, {Sooley},
  {Sornarajah}, {Streicher}, {Teuben}, {Thomas}, {Tremblay}, {Turner},
  {Terr{\'o}n}, {van Kerkwijk}, {de la Vega}, {Watkins}, {Weaver}, {Whitmore},
  {Woillez}, {Zabalza}, \& {Astropy Contributors}}]{astropy:2018}
{Astropy Collaboration}, {Price-Whelan}, A.~M., {Sip{\H{o}}cz}, B.~M., {et~al.}
  2018, \aj, 156, 123, \dodoi{10.3847/1538-3881/aabc4f}

\bibitem[{{Bellm} {et~al.}(2019){Bellm}, {Kulkarni}, {Graham}, {Dekany},
  {Smith}, {Riddle}, {Masci}, {Helou}, {Prince}, {Adams}, {Barbarino},
  {Barlow}, {Bauer}, {Beck}, {Belicki}, {Biswas}, {Blagorodnova}, {Bodewits},
  {Bolin}, {Brinnel}, {Brooke}, {Bue}, {Bulla}, {Burruss}, {Cenko}, {Chang},
  {Connolly}, {Coughlin}, {Cromer}, {Cunningham}, {De}, {Delacroix}, {Desai},
  {Duev}, {Eadie}, {Farnham}, {Feeney}, {Feindt}, {Flynn}, {Franckowiak},
  {Frederick}, {Fremling}, {Gal-Yam}, {Gezari}, {Giomi}, {Goldstein},
  {Golkhou}, {Goobar}, {Groom}, {Hacopians}, {Hale}, {Henning}, {Ho}, {Hover},
  {Howell}, {Hung}, {Huppenkothen}, {Imel}, {Ip}, {Ivezi{\'c}}, {Jackson},
  {Jones}, {Juric}, {Kasliwal}, {Kaspi}, {Kaye}, {Kelley}, {Kowalski},
  {Kramer}, {Kupfer}, {Landry}, {Laher}, {Lee}, {Lin}, {Lin}, {Lunnan},
  {Giomi}, {Mahabal}, {Mao}, {Miller}, {Monkewitz}, {Murphy}, {Ngeow},
  {Nordin}, {Nugent}, {Ofek}, {Patterson}, {Penprase}, {Porter}, {Rauch},
  {Rebbapragada}, {Reiley}, {Rigault}, {Rodriguez}, {van Roestel}, {Rusholme},
  {van Santen}, {Schulze}, {Shupe}, {Singer}, {Soumagnac}, {Stein}, {Surace},
  {Sollerman}, {Szkody}, {Taddia}, {Terek}, {Van Sistine}, {van Velzen},
  {Vestrand}, {Walters}, {Ward}, {Ye}, {Yu}, {Yan}, \&
  {Zolkower}}]{2019PASP..131a8002B}
{Bellm}, E.~C., {Kulkarni}, S.~R., {Graham}, M.~J., {et~al.} 2019, \pasp, 131,
  018002, \dodoi{10.1088/1538-3873/aaecbe}

\bibitem[{{Belokurov} {et~al.}(2018){Belokurov}, {Erkal}, {Evans}, {Koposov},
  \& {Deason}}]{2018MNRAS.478..611B}
{Belokurov}, V., {Erkal}, D., {Evans}, N.~W., {Koposov}, S.~E., \& {Deason},
  A.~J. 2018, \mnras, 478, 611, \dodoi{10.1093/mnras/sty982}

\bibitem[{{Bensby} {et~al.}(2003){Bensby}, {Feltzing}, \&
  {Lundstr{\"o}m}}]{2003A&A...410..527B}
{Bensby}, T., {Feltzing}, S., \& {Lundstr{\"o}m}, I. 2003, \aap, 410, 527,
  \dodoi{10.1051/0004-6361:20031213}

\bibitem[{{Bovy}(2015)}]{2015ApJS..216...29B}
{Bovy}, J. 2015, \apjs, 216, 29, \dodoi{10.1088/0067-0049/216/2/29}

\bibitem[{{Carollo} {et~al.}(2007){Carollo}, {Beers}, {Lee}, {Chiba}, {Norris},
  {Wilhelm}, {Sivarani}, {Marsteller}, {Munn}, {Bailer-Jones}, {Fiorentin}, \&
  {York}}]{2007Natur.450.1020C}
{Carollo}, D., {Beers}, T.~C., {Lee}, Y.~S., {et~al.} 2007, \nat, 450, 1020,
  \dodoi{10.1038/nature06460}

\bibitem[{{Cropper} {et~al.}(2018){Cropper}, {Katz}, {Sartoretti}, {Prusti},
  {de Bruijne}, {Chassat}, {Charvet}, {Boyadjian}, {Perryman}, {Sarri}, {Gare},
  {Erdmann}, {Munari}, {Zwitter}, {Wilkinson}, {Arenou}, {Vallenari},
  {G{\'o}mez}, {Panuzzo}, {Seabroke}, {Allende Prieto}, {Benson}, {Marchal},
  {Huckle}, {Smith}, {Dolding}, {Jan{\ss}en}, {Viala}, {Blomme}, {Baker},
  {Boudreault}, {Crifo}, {Soubiran}, {Fr{\'e}mat}, {Jasniewicz}, {Guerrier},
  {Guy}, {Turon}, {Jean-Antoine-Piccolo}, {Th{\'e}venin}, {David}, {Gosset}, \&
  {Damerdji}}]{2018A&A...616A...5C}
{Cropper}, M., {Katz}, D., {Sartoretti}, P., {et~al.} 2018, \aap, 616, A5,
  \dodoi{10.1051/0004-6361/201832763}

\bibitem[{{Dark Energy Survey Collaboration} {et~al.}(2016){Dark Energy Survey
  Collaboration}, {Abbott}, {Abdalla}, {Aleksi{\'c}}, {Allam}, {Amara},
  {Bacon}, {Balbinot}, {Banerji}, {Bechtol}, {Benoit-L{\'e}vy}, {Bernstein},
  {Bertin}, {Blazek}, {Bonnett}, {Bridle}, {Brooks}, {Brunner}, {Buckley-Geer},
  {Burke}, {Caminha}, {Capozzi}, {Carlsen}, {Carnero-Rosell}, {Carollo},
  {Carrasco-Kind}, {Carretero}, {Castander}, {Clerkin}, {Collett}, {Conselice},
  {Crocce}, {Cunha}, {D'Andrea}, {da Costa}, {Davis}, {Desai}, {Diehl},
  {Dietrich}, {Dodelson}, {Doel}, {Drlica-Wagner}, {Estrada}, {Etherington},
  {Evrard}, {Fabbri}, {Finley}, {Flaugher}, {Foley}, {Fosalba}, {Frieman},
  {Garc{\'\i}a-Bellido}, {Gaztanaga}, {Gerdes}, {Giannantonio}, {Goldstein},
  {Gruen}, {Gruendl}, {Guarnieri}, {Gutierrez}, {Hartley}, {Honscheid}, {Jain},
  {James}, {Jeltema}, {Jouvel}, {Kessler}, {King}, {Kirk}, {Kron}, {Kuehn},
  {Kuropatkin}, {Lahav}, {Li}, {Lima}, {Lin}, {Maia}, {Makler}, {Manera},
  {Maraston}, {Marshall}, {Martini}, {McMahon}, {Melchior}, {Merson}, {Miller},
  {Miquel}, {Mohr}, {Morice-Atkinson}, {Naidoo}, {Neilsen}, {Nichol}, {Nord},
  {Ogando}, {Ostrovski}, {Palmese}, {Papadopoulos}, {Peiris}, {Peoples},
  {Percival}, {Plazas}, {Reed}, {Refregier}, {Romer}, {Roodman}, {Ross},
  {Rozo}, {Rykoff}, {Sadeh}, {Sako}, {S{\'a}nchez}, {Sanchez}, {Santiago},
  {Scarpine}, {Schubnell}, {Sevilla-Noarbe}, {Sheldon}, {Smith}, {Smith},
  {Soares-Santos}, {Sobreira}, {Soumagnac}, {Suchyta}, {Sullivan}, {Swanson},
  {Tarle}, {Thaler}, {Thomas}, {Thomas}, {Tucker}, {Vieira}, {Vikram},
  {Walker}, {Wechsler}, {Weller}, {Wester}, {Whiteway}, {Wilcox}, {Yanny},
  {Zhang}, \& {Zuntz}}]{2016MNRAS.460.1270D}
{Dark Energy Survey Collaboration}, {Abbott}, T., {Abdalla}, F.~B., {et~al.}
  2016, \mnras, 460, 1270, \dodoi{10.1093/mnras/stw641}

\bibitem[{{Dekker} {et~al.}(2000){Dekker}, {D'Odorico}, {Kaufer}, {Delabre}, \&
  {Kotzlowski}}]{2000SPIE.4008..534D}
{Dekker}, H., {D'Odorico}, S., {Kaufer}, A., {Delabre}, B., \& {Kotzlowski}, H.
  2000, in Society of Photo-Optical Instrumentation Engineers (SPIE) Conference
  Series, Vol. 4008, Optical and IR Telescope Instrumentation and Detectors,
  ed. M.~{Iye} \& A.~F. {Moorwood}, 534--545, \dodoi{10.1117/12.395512}

\bibitem[{{Eggen}(1958)}]{1958MNRAS.118...65E}
{Eggen}, O.~J. 1958, \mnras, 118, 65, \dodoi{10.1093/mnras/118.1.65}

\bibitem[{Foreman-Mackey(2016)}]{corner}
Foreman-Mackey, D. 2016, The Journal of Open Source Software, 1, 24,
  \dodoi{10.21105/joss.00024}

\bibitem[{{Foreman-Mackey} {et~al.}(2013){Foreman-Mackey}, {Hogg}, {Lang}, \&
  {Goodman}}]{2013PASP..125..306F}
{Foreman-Mackey}, D., {Hogg}, D.~W., {Lang}, D., \& {Goodman}, J. 2013, \pasp,
  125, 306, \dodoi{10.1086/670067}

\bibitem[{{Freeman}(1993)}]{1993ASPC...48..608F}
{Freeman}, K.~C. 1993, in Astronomical Society of the Pacific Conference
  Series, Vol.~48, The Globular Cluster-Galaxy Connection, ed. G.~H. {Smith} \&
  J.~P. {Brodie}, 608

\bibitem[{{Freudling} {et~al.}(2013){Freudling}, {Romaniello}, {Bramich},
  {Ballester}, {Forchi}, {Garc{\'\i}a-Dabl{\'o}}, {Moehler}, \&
  {Neeser}}]{2013A&A...559A..96F}
{Freudling}, W., {Romaniello}, M., {Bramich}, D.~M., {et~al.} 2013, \aap, 559,
  A96, \dodoi{10.1051/0004-6361/201322494}

\bibitem[{{Gaia Collaboration} {et~al.}(2018{\natexlab{a}}){Gaia
  Collaboration}, {Brown}, {Vallenari}, {Prusti}, {de Bruijne}, {Babusiaux},
  {Bailer-Jones}, {Biermann}, {Evans}, {Eyer}, {Jansen}, {Jordi}, {Klioner},
  {Lammers}, {Lindegren}, {Luri}, {Mignard}, {Panem}, {Pourbaix}, {Randich},
  {Sartoretti}, {Siddiqui}, {Soubiran}, {van Leeuwen}, {Walton}, {Arenou},
  {Bastian}, {Cropper}, {Drimmel}, {Katz}, {Lattanzi}, {Bakker}, {Cacciari},
  {Casta{\~n}eda}, {Chaoul}, {Cheek}, {De Angeli}, {Fabricius}, {Guerra},
  {Holl}, {Masana}, {Messineo}, {Mowlavi}, {Nienartowicz}, {Panuzzo},
  {Portell}, {Riello}, {Seabroke}, {Tanga}, {Th{\'e}venin}, {Gracia-Abril},
  {Comoretto}, {Garcia-Reinaldos}, {Teyssier}, {Altmann}, {Andrae}, {Audard},
  {Bellas-Velidis}, {Benson}, {Berthier}, {Blomme}, {Burgess}, {Busso},
  {Carry}, {Cellino}, {Clementini}, {Clotet}, {Creevey}, {Davidson}, {De
  Ridder}, {Delchambre}, {Dell'Oro}, {Ducourant},
  {Fern{\'a}ndez-Hern{\'a}ndez}, {Fouesneau}, {Fr{\'e}mat}, {Galluccio},
  {Garc{\'\i}a-Torres}, {Gonz{\'a}lez-N{\'u}{\~n}ez}, {Gonz{\'a}lez-Vidal},
  {Gosset}, {Guy}, {Halbwachs}, {Hambly}, {Harrison}, {Hern{\'a}ndez},
  {Hestroffer}, {Hodgkin}, {Hutton}, {Jasniewicz}, {Jean-Antoine-Piccolo},
  {Jordan}, {Korn}, {Krone-Martins}, {Lanzafame}, {Lebzelter}, {L{\"o}ffler},
  {Manteiga}, {Marrese}, {Mart{\'\i}n-Fleitas}, {Moitinho}, {Mora}, {Muinonen},
  {Osinde}, {Pancino}, {Pauwels}, {Petit}, {Recio-Blanco}, {Richards},
  {Rimoldini}, {Robin}, {Sarro}, {Siopis}, {Smith}, {Sozzetti}, {S{\"u}veges},
  {Torra}, {van Reeven}, {Abbas}, {Abreu Aramburu}, {Accart}, {Aerts},
  {Altavilla}, {{\'A}lvarez}, {Alvarez}, {Alves}, {Anderson}, {Andrei},
  {Anglada Varela}, {Antiche}, {Antoja}, {Arcay}, {Astraatmadja}, {Bach},
  {Baker}, {Balaguer-N{\'u}{\~n}ez}, {Balm}, {Barache}, {Barata}, {Barbato},
  {Barblan}, {Barklem}, {Barrado}, {Barros}, {Barstow}, {Bartholom{\'e}
  Mu{\~n}oz}, {Bassilana}, {Becciani}, {Bellazzini}, {Berihuete}, {Bertone},
  {Bianchi}, {Bienaym{\'e}}, {Blanco-Cuaresma}, {Boch}, {Boeche}, {Bombrun},
  {Borrachero}, {Bossini}, {Bouquillon}, {Bourda}, {Bragaglia}, {Bramante},
  {Breddels}, {Bressan}, {Brouillet}, {Br{\"u}semeister}, {Brugaletta},
  {Bucciarelli}, {Burlacu}, {Busonero}, {Butkevich}, {Buzzi}, {Caffau},
  {Cancelliere}, {Cannizzaro}, {Cantat-Gaudin}, {Carballo}, {Carlucci},
  {Carrasco}, {Casamiquela}, {Castellani}, {Castro-Ginard}, {Charlot},
  {Chemin}, {Chiavassa}, {Cocozza}, {Costigan}, {Cowell}, {Crifo}, {Crosta},
  {Crowley}, {Cuypers}, {Dafonte}, {Damerdji}, {Dapergolas}, {David}, {David},
  {de Laverny}, {De Luise}, {De March}, {de Martino}, {de Souza}, {de Torres},
  {Debosscher}, {del Pozo}, {Delbo}, {Delgado}, {Delgado}, {Di Matteo},
  {Diakite}, {Diener}, {Distefano}, {Dolding}, {Drazinos}, {Dur{\'a}n},
  {Edvardsson}, {Enke}, {Eriksson}, {Esquej}, {Eynard Bontemps}, {Fabre},
  {Fabrizio}, {Faigler}, {Falc{\~a}o}, {Farr{\`a}s Casas}, {Federici},
  {Fedorets}, {Fernique}, {Figueras}, {Filippi}, {Findeisen}, {Fonti},
  {Fraile}, {Fraser}, {Fr{\'e}zouls}, {Gai}, {Galleti}, {Garabato},
  {Garc{\'\i}a-Sedano}, {Garofalo}, {Garralda}, {Gavel}, {Gavras}, {Gerssen},
  {Geyer}, {Giacobbe}, {Gilmore}, {Girona}, {Giuffrida}, {Glass}, {Gomes},
  {Granvik}, {Gueguen}, {Guerrier}, {Guiraud}, {Guti{\'e}rrez-S{\'a}nchez},
  {Haigron}, {Hatzidimitriou}, {Hauser}, {Haywood}, {Heiter}, {Helmi}, {Heu},
  {Hilger}, {Hobbs}, {Hofmann}, {Holland}, {Huckle}, {Hypki}, {Icardi},
  {Jan{\ss}en}, {Jevardat de Fombelle}, {Jonker}, {Juh{\'a}sz}, {Julbe},
  {Karampelas}, {Kewley}, {Klar}, {Kochoska}, {Kohley}, {Kolenberg},
  {Kontizas}, {Kontizas}, {Koposov}, {Kordopatis}, {Kostrzewa-Rutkowska},
  {Koubsky}, {Lambert}, {Lanza}, {Lasne}, {Lavigne}, {Le Fustec}, {Le
  Poncin-Lafitte}, {Lebreton}, {Leccia}, {Leclerc}, {Lecoeur-Taibi},
  {Lenhardt}, {Leroux}, {Liao}, {Licata}, {Lindstr{\o}m}, {Lister}, {Livanou},
  {Lobel}, {L{\'o}pez}, {Managau}, {Mann}, {Mantelet}, {Marchal}, {Marchant},
  {Marconi}, {Marinoni}, {Marschalk{\'o}}, {Marshall}, {Martino}, {Marton},
  {Mary}, {Massari}, {Matijevi{\v{c}}}, {Mazeh}, {McMillan}, {Messina},
  {Michalik}, {Millar}, {Molina}, {Molinaro}, {Moln{\'a}r}, {Montegriffo},
  {Mor}, {Morbidelli}, {Morel}, {Morris}, {Mulone}, {Muraveva}, {Musella},
  {Nelemans}, {Nicastro}, {Noval}, {O'Mullane}, {Ord{\'e}novic},
  {Ord{\'o}{\~n}ez-Blanco}, {Osborne}, {Pagani}, {Pagano}, {Pailler},
  {Palacin}, {Palaversa}, {Panahi}, {Pawlak}, {Piersimoni}, {Pineau}, {Plachy},
  {Plum}, {Poggio}, {Poujoulet}, {Pr{\v{s}}a}, {Pulone}, {Racero}, {Ragaini},
  {Rambaux}, {Ramos-Lerate}, {Regibo}, {Reyl{\'e}}, {Riclet}, {Ripepi}, {Riva},
  {Rivard}, {Rixon}, {Roegiers}, {Roelens}, {Romero-G{\'o}mez}, {Rowell},
  {Royer}, {Ruiz-Dern}, {Sadowski}, {Sagrist{\`a} Sell{\'e}s}, {Sahlmann},
  {Salgado}, {Salguero}, {Sanna}, {Santana-Ros}, {Sarasso}, {Savietto},
  {Schultheis}, {Sciacca}, {Segol}, {Segovia}, {S{\'e}gransan}, {Shih},
  {Siltala}, {Silva}, {Smart}, {Smith}, {Solano}, {Solitro}, {Sordo}, {Soria
  Nieto}, {Souchay}, {Spagna}, {Spoto}, {Stampa}, {Steele},
  {Steidelm{\"u}ller}, {Stephenson}, {Stoev}, {Suess}, {Surdej}, {Szabados},
  {Szegedi-Elek}, {Tapiador}, {Taris}, {Tauran}, {Taylor}, {Teixeira},
  {Terrett}, {Teyssandier}, {Thuillot}, {Titarenko}, {Torra Clotet}, {Turon},
  {Ulla}, {Utrilla}, {Uzzi}, {Vaillant}, {Valentini}, {Valette}, {van Elteren},
  {Van Hemelryck}, {van Leeuwen}, {Vaschetto}, {Vecchiato}, {Veljanoski},
  {Viala}, {Vicente}, {Vogt}, {von Essen}, {Voss}, {Votruba}, {Voutsinas},
  {Walmsley}, {Weiler}, {Wertz}, {Wevers}, {Wyrzykowski}, {Yoldas},
  {{\v{Z}}erjal}, {Ziaeepour}, {Zorec}, {Zschocke}, {Zucker}, {Zurbach}, \&
  {Zwitter}}]{2018A&A...616A...1G}
{Gaia Collaboration}, {Brown}, A.~G.~A., {Vallenari}, A., {et~al.}
  2018{\natexlab{a}}, \aap, 616, A1, \dodoi{10.1051/0004-6361/201833051}

\bibitem[{{Gaia Collaboration} {et~al.}(2018{\natexlab{b}}){Gaia
  Collaboration}, {Babusiaux}, {van Leeuwen}, {Barstow}, {Jordi}, {Vallenari},
  {Bossini}, {Bressan}, {Cantat-Gaudin}, {van Leeuwen}, {Brown}, {Prusti}, {de
  Bruijne}, {Bailer-Jones}, {Biermann}, {Evans}, {Eyer}, {Jansen}, {Klioner},
  {Lammers}, {Lindegren}, {Luri}, {Mignard}, {Panem}, {Pourbaix}, {Randich},
  {Sartoretti}, {Siddiqui}, {Soubiran}, {Walton}, {Arenou}, {Bastian},
  {Cropper}, {Drimmel}, {Katz}, {Lattanzi}, {Bakker}, {Cacciari},
  {Casta{\~n}eda}, {Chaoul}, {Cheek}, {De Angeli}, {Fabricius}, {Guerra},
  {Holl}, {Masana}, {Messineo}, {Mowlavi}, {Nienartowicz}, {Panuzzo},
  {Portell}, {Riello}, {Seabroke}, {Tanga}, {Th{\'e}venin}, {Gracia-Abril},
  {Comoretto}, {Garcia-Reinaldos}, {Teyssier}, {Altmann}, {Andrae}, {Audard},
  {Bellas-Velidis}, {Benson}, {Berthier}, {Blomme}, {Burgess}, {Busso},
  {Carry}, {Cellino}, {Clementini}, {Clotet}, {Creevey}, {Davidson}, {De
  Ridder}, {Delchambre}, {Dell'Oro}, {Ducourant},
  {Fern{\'a}ndez-Hern{\'a}ndez}, {Fouesneau}, {Fr{\'e}mat}, {Galluccio},
  {Garc{\'\i}a-Torres}, {Gonz{\'a}lez-N{\'u}{\~n}ez}, {Gonz{\'a}lez-Vidal},
  {Gosset}, {Guy}, {Halbwachs}, {Hambly}, {Harrison}, {Hern{\'a}ndez},
  {Hestroffer}, {Hodgkin}, {Hutton}, {Jasniewicz}, {Jean-Antoine-Piccolo},
  {Jordan}, {Korn}, {Krone-Martins}, {Lanzafame}, {Lebzelter}, {L{\"o}ffler},
  {Manteiga}, {Marrese}, {Mart{\'\i}n-Fleitas}, {Moitinho}, {Mora}, {Muinonen},
  {Osinde}, {Pancino}, {Pauwels}, {Petit}, {Recio-Blanco}, {Richards},
  {Rimoldini}, {Robin}, {Sarro}, {Siopis}, {Smith}, {Sozzetti}, {S{\"u}veges},
  {Torra}, {van Reeven}, {Abbas}, {Abreu Aramburu}, {Accart}, {Aerts},
  {Altavilla}, {{\'A}lvarez}, {Alvarez}, {Alves}, {Anderson}, {Andrei},
  {Anglada Varela}, {Antiche}, {Antoja}, {Arcay}, {Astraatmadja}, {Bach},
  {Baker}, {Balaguer-N{\'u}{\~n}ez}, {Balm}, {Barache}, {Barata}, {Barbato},
  {Barblan}, {Barklem}, {Barrado}, {Barros}, {Bartholom{\'e} Mu{\~n}oz},
  {Bassilana}, {Becciani}, {Bellazzini}, {Berihuete}, {Bertone}, {Bianchi},
  {Bienaym{\'e}}, {Blanco-Cuaresma}, {Boch}, {Boeche}, {Bombrun}, {Borrachero},
  {Bouquillon}, {Bourda}, {Bragaglia}, {Bramante}, {Breddels}, {Brouillet},
  {Br{\"u}semeister}, {Brugaletta}, {Bucciarelli}, {Burlacu}, {Busonero},
  {Butkevich}, {Buzzi}, {Caffau}, {Cancelliere}, {Cannizzaro}, {Carballo},
  {Carlucci}, {Carrasco}, {Casamiquela}, {Castellani}, {Castro-Ginard},
  {Charlot}, {Chemin}, {Chiavassa}, {Cocozza}, {Costigan}, {Cowell}, {Crifo},
  {Crosta}, {Crowley}, {Cuypers}, {Dafonte}, {Damerdji}, {Dapergolas}, {David},
  {David}, {de Laverny}, {De Luise}, {De March}, {de Martino}, {de Souza}, {de
  Torres}, {Debosscher}, {del Pozo}, {Delbo}, {Delgado}, {Delgado}, {Diakite},
  {Diener}, {Distefano}, {Dolding}, {Drazinos}, {Dur{\'a}n}, {Edvardsson},
  {Enke}, {Eriksson}, {Esquej}, {Eynard Bontemps}, {Fabre}, {Fabrizio},
  {Faigler}, {Falc{\~a}o}, {Farr{\`a}s Casas}, {Federici}, {Fedorets},
  {Fernique}, {Figueras}, {Filippi}, {Findeisen}, {Fonti}, {Fraile}, {Fraser},
  {Fr{\'e}zouls}, {Gai}, {Galleti}, {Garabato}, {Garc{\'\i}a-Sedano},
  {Garofalo}, {Garralda}, {Gavel}, {Gavras}, {Gerssen}, {Geyer}, {Giacobbe},
  {Gilmore}, {Girona}, {Giuffrida}, {Glass}, {Gomes}, {Granvik}, {Gueguen},
  {Guerrier}, {Guiraud}, {Guti{\'e}}, {Haigron}, {Hatzidimitriou}, {Hauser},
  {Haywood}, {Heiter}, {Helmi}, {Heu}, {Hilger}, {Hobbs}, {Hofmann}, {Holland},
  {Huckle}, {Hypki}, {Icardi}, {Jan{\ss}en}, {Jevardat de Fombelle}, {Jonker},
  {Juh{\'a}sz}, {Julbe}, {Karampelas}, {Kewley}, {Klar}, {Kochoska}, {Kohley},
  {Kolenberg}, {Kontizas}, {Kontizas}, {Koposov}, {Kordopatis},
  {Kostrzewa-Rutkowska}, {Koubsky}, {Lambert}, {Lanza}, {Lasne}, {Lavigne}, {Le
  Fustec}, {Le Poncin-Lafitte}, {Lebreton}, {Leccia}, {Leclerc},
  {Lecoeur-Taibi}, {Lenhardt}, {Leroux}, {Liao}, {Licata}, {Lindstr{\o}m},
  {Lister}, {Livanou}, {Lobel}, {L{\'o}pez}, {Managau}, {Mann}, {Mantelet},
  {Marchal}, {Marchant}, {Marconi}, {Marinoni}, {Marschalk{\'o}}, {Marshall},
  {Martino}, {Marton}, {Mary}, {Massari}, {Matijevi{\v{c}}}, {Mazeh},
  {McMillan}, {Messina}, {Michalik}, {Millar}, {Molina}, {Molinaro},
  {Moln{\'a}r}, {Montegriffo}, {Mor}, {Morbidelli}, {Morel}, {Morris},
  {Mulone}, {Muraveva}, {Musella}, {Nelemans}, {Nicastro}, {Noval},
  {O'Mullane}, {Ord{\'e}novic}, {Ord{\'o}{\~n}ez-Blanco}, {Osborne}, {Pagani},
  {Pagano}, {Pailler}, {Palacin}, {Palaversa}, {Panahi}, {Pawlak},
  {Piersimoni}, {Pineau}, {Plachy}, {Plum}, {Poggio}, {Poujoulet},
  {Pr{\v{s}}a}, {Pulone}, {Racero}, {Ragaini}, {Rambaux}, {Ramos-Lerate},
  {Regibo}, {Reyl{\'e}}, {Riclet}, {Ripepi}, {Riva}, {Rivard}, {Rixon},
  {Roegiers}, {Roelens}, {Romero-G{\'o}mez}, {Rowell}, {Royer}, {Ruiz-Dern},
  {Sadowski}, {Sagrist{\`a} Sell{\'e}s}, {Sahlmann}, {Salgado}, {Salguero},
  {Sanna}, {Santana-Ros}, {Sarasso}, {Savietto}, {Schultheis}, {Sciacca},
  {Segol}, {Segovia}, {S{\'e}gransan}, {Shih}, {Siltala}, {Silva}, {Smart},
  {Smith}, {Solano}, {Solitro}, {Sordo}, {Soria Nieto}, {Souchay}, {Spagna},
  {Spoto}, {Stampa}, {Steele}, {Steidelm{\"u}ller}, {Stephenson}, {Stoev},
  {Suess}, {Surdej}, {Szabados}, {Szegedi-Elek}, {Tapiador}, {Taris}, {Tauran},
  {Taylor}, {Teixeira}, {Terrett}, {Teyssandier}, {Thuillot}, {Titarenko},
  {Torra Clotet}, {Turon}, {Ulla}, {Utrilla}, {Uzzi}, {Vaillant}, {Valentini},
  {Valette}, {van Elteren}, {Van Hemelryck}, {Vaschetto}, {Vecchiato},
  {Veljanoski}, {Viala}, {Vicente}, {Vogt}, {von Essen}, {Voss}, {Votruba},
  {Voutsinas}, {Walmsley}, {Weiler}, {Wertz}, {Wevers}, {Wyrzykowski},
  {Yoldas}, {{\v{Z}}erjal}, {Ziaeepour}, {Zorec}, {Zschocke}, {Zucker},
  {Zurbach}, \& {Zwitter}}]{2018A&A...616A..10G}
{Gaia Collaboration}, {Babusiaux}, C., {van Leeuwen}, F., {et~al.}
  2018{\natexlab{b}}, \aap, 616, A10, \dodoi{10.1051/0004-6361/201832843}

\bibitem[{{Gaia Collaboration} {et~al.}(2021){Gaia Collaboration}, {Brown},
  {Vallenari}, {Prusti}, {de Bruijne}, {Babusiaux}, {Biermann}, {Creevey},
  {Evans}, {Eyer}, {Hutton}, {Jansen}, {Jordi}, {Klioner}, {Lammers},
  {Lindegren}, {Luri}, {Mignard}, {Panem}, {Pourbaix}, {Randich}, {Sartoretti},
  {Soubiran}, {Walton}, {Arenou}, {Bailer-Jones}, {Bastian}, {Cropper},
  {Drimmel}, {Katz}, {Lattanzi}, {van Leeuwen}, {Bakker}, {Cacciari},
  {Casta{\~n}eda}, {De Angeli}, {Ducourant}, {Fabricius}, {Fouesneau},
  {Fr{\'e}mat}, {Guerra}, {Guerrier}, {Guiraud}, {Jean-Antoine Piccolo},
  {Masana}, {Messineo}, {Mowlavi}, {Nicolas}, {Nienartowicz}, {Pailler},
  {Panuzzo}, {Riclet}, {Roux}, {Seabroke}, {Sordo}, {Tanga}, {Th{\'e}venin},
  {Gracia-Abril}, {Portell}, {Teyssier}, {Altmann}, {Andrae}, {Bellas-Velidis},
  {Benson}, {Berthier}, {Blomme}, {Brugaletta}, {Burgess}, {Busso}, {Carry},
  {Cellino}, {Cheek}, {Clementini}, {Damerdji}, {Davidson}, {Delchambre},
  {Dell'Oro}, {Fern{\'a}ndez-Hern{\'a}ndez}, {Galluccio}, {Garc{\'\i}a-Lario},
  {Garcia-Reinaldos}, {Gonz{\'a}lez-N{\'u}{\~n}ez}, {Gosset}, {Haigron},
  {Halbwachs}, {Hambly}, {Harrison}, {Hatzidimitriou}, {Heiter},
  {Hern{\'a}ndez}, {Hestroffer}, {Hodgkin}, {Holl}, {Jan{\ss}en}, {Jevardat de
  Fombelle}, {Jordan}, {Krone-Martins}, {Lanzafame}, {L{\"o}ffler}, {Lorca},
  {Manteiga}, {Marchal}, {Marrese}, {Moitinho}, {Mora}, {Muinonen}, {Osborne},
  {Pancino}, {Pauwels}, {Petit}, {Recio-Blanco}, {Richards}, {Riello},
  {Rimoldini}, {Robin}, {Roegiers}, {Rybizki}, {Sarro}, {Siopis}, {Smith},
  {Sozzetti}, {Ulla}, {Utrilla}, {van Leeuwen}, {van Reeven}, {Abbas}, {Abreu
  Aramburu}, {Accart}, {Aerts}, {Aguado}, {Ajaj}, {Altavilla}, {{\'A}lvarez},
  {{\'A}lvarez Cid-Fuentes}, {Alves}, {Anderson}, {Anglada Varela}, {Antoja},
  {Audard}, {Baines}, {Baker}, {Balaguer-N{\'u}{\~n}ez}, {Balbinot}, {Balog},
  {Barache}, {Barbato}, {Barros}, {Barstow}, {Bartolom{\'e}}, {Bassilana},
  {Bauchet}, {Baudesson-Stella}, {Becciani}, {Bellazzini}, {Bernet}, {Bertone},
  {Bianchi}, {Blanco-Cuaresma}, {Boch}, {Bombrun}, {Bossini}, {Bouquillon},
  {Bragaglia}, {Bramante}, {Breedt}, {Bressan}, {Brouillet}, {Bucciarelli},
  {Burlacu}, {Busonero}, {Butkevich}, {Buzzi}, {Caffau}, {Cancelliere},
  {C{\'a}novas}, {Cantat-Gaudin}, {Carballo}, {Carlucci}, {Carnerero},
  {Carrasco}, {Casamiquela}, {Castellani}, {Castro-Ginard}, {Castro Sampol},
  {Chaoul}, {Charlot}, {Chemin}, {Chiavassa}, {Cioni}, {Comoretto}, {Cooper},
  {Cornez}, {Cowell}, {Crifo}, {Crosta}, {Crowley}, {Dafonte}, {Dapergolas},
  {David}, {David}, {de Laverny}, {De Luise}, {De March}, {De Ridder}, {de
  Souza}, {de Teodoro}, {de Torres}, {del Peloso}, {del Pozo}, {Delbo},
  {Delgado}, {Delgado}, {Delisle}, {Di Matteo}, {Diakite}, {Diener},
  {Distefano}, {Dolding}, {Eappachen}, {Edvardsson}, {Enke}, {Esquej}, {Fabre},
  {Fabrizio}, {Faigler}, {Fedorets}, {Fernique}, {Fienga}, {Figueras},
  {Fouron}, {Fragkoudi}, {Fraile}, {Franke}, {Gai}, {Garabato},
  {Garcia-Gutierrez}, {Garc{\'\i}a-Torres}, {Garofalo}, {Gavras}, {Gerlach},
  {Geyer}, {Giacobbe}, {Gilmore}, {Girona}, {Giuffrida}, {Gomel}, {Gomez},
  {Gonzalez-Santamaria}, {Gonz{\'a}lez-Vidal}, {Granvik},
  {Guti{\'e}rrez-S{\'a}nchez}, {Guy}, {Hauser}, {Haywood}, {Helmi}, {Hidalgo},
  {Hilger}, {H{\l}adczuk}, {Hobbs}, {Holland}, {Huckle}, {Jasniewicz},
  {Jonker}, {Juaristi Campillo}, {Julbe}, {Karbevska}, {Kervella}, {Khanna},
  {Kochoska}, {Kontizas}, {Kordopatis}, {Korn}, {Kostrzewa-Rutkowska},
  {Kruszy{\'n}ska}, {Lambert}, {Lanza}, {Lasne}, {Le Campion}, {Le Fustec},
  {Lebreton}, {Lebzelter}, {Leccia}, {Leclerc}, {Lecoeur-Taibi}, {Liao},
  {Licata}, {Lindstr{\o}m}, {Lister}, {Livanou}, {Lobel}, {Madrero Pardo},
  {Managau}, {Mann}, {Marchant}, {Marconi}, {Marcos Santos}, {Marinoni},
  {Marocco}, {Marshall}, {Martin Polo}, {Mart{\'\i}n-Fleitas}, {Masip},
  {Massari}, {Mastrobuono-Battisti}, {Mazeh}, {McMillan}, {Messina},
  {Michalik}, {Millar}, {Mints}, {Molina}, {Molinaro}, {Moln{\'a}r},
  {Montegriffo}, {Mor}, {Morbidelli}, {Morel}, {Morris}, {Mulone}, {Munoz},
  {Muraveva}, {Murphy}, {Musella}, {Noval}, {Ord{\'e}novic}, {Orr{\`u}},
  {Osinde}, {Pagani}, {Pagano}, {Palaversa}, {Palicio}, {Panahi}, {Pawlak},
  {Pe{\~n}alosa Esteller}, {Penttil{\"a}}, {Piersimoni}, {Pineau}, {Plachy},
  {Plum}, {Poggio}, {Poretti}, {Poujoulet}, {Pr{\v{s}}a}, {Pulone}, {Racero},
  {Ragaini}, {Rainer}, {Raiteri}, {Rambaux}, {Ramos}, {Ramos-Lerate}, {Re
  Fiorentin}, {Regibo}, {Reyl{\'e}}, {Ripepi}, {Riva}, {Rixon}, {Robichon},
  {Robin}, {Roelens}, {Rohrbasser}, {Romero-G{\'o}mez}, {Rowell}, {Royer},
  {Rybicki}, {Sadowski}, {Sagrist{\`a} Sell{\'e}s}, {Sahlmann}, {Salgado},
  {Salguero}, {Samaras}, {Sanchez Gimenez}, {Sanna}, {Santove{\~n}a},
  {Sarasso}, {Schultheis}, {Sciacca}, {Segol}, {Segovia}, {S{\'e}gransan},
  {Semeux}, {Shahaf}, {Siddiqui}, {Siebert}, {Siltala}, {Slezak}, {Smart},
  {Solano}, {Solitro}, {Souami}, {Souchay}, {Spagna}, {Spoto}, {Steele},
  {Steidelm{\"u}ller}, {Stephenson}, {S{\"u}veges}, {Szabados}, {Szegedi-Elek},
  {Taris}, {Tauran}, {Taylor}, {Teixeira}, {Thuillot}, {Tonello}, {Torra},
  {Torra}, {Turon}, {Unger}, {Vaillant}, {van Dillen}, {Vanel}, {Vecchiato},
  {Viala}, {Vicente}, {Voutsinas}, {Weiler}, {Wevers}, {Wyrzykowski}, {Yoldas},
  {Yvard}, {Zhao}, {Zorec}, {Zucker}, {Zurbach}, \&
  {Zwitter}}]{2021A&A...649A...1G}
{Gaia Collaboration}, {Brown}, A.~G.~A., {Vallenari}, A., {et~al.} 2021, \aap,
  649, A1, \dodoi{10.1051/0004-6361/202039657}

\bibitem[{{Giclas} {et~al.}(1975){Giclas}, {Burnham}, \&
  {Thomas}}]{1975LowOB...8....9G}
{Giclas}, H.~L., {Burnham}, R., J., \& {Thomas}, N.~G. 1975, Lowell Observatory
  Bulletin, 8, 9

\bibitem[{{Giclas} {et~al.}(1961){Giclas}, {Burnham}, \&
  {Thomas}}]{1961LowOB...5...61G}
{Giclas}, H.~L., {Burnham}, R., \& {Thomas}, N.~G. 1961, Lowell Observatory
  Bulletin, 6, 61

\bibitem[{{Giclas} {et~al.}(1959){Giclas}, {Slaughter}, \&
  {Burnham}}]{1959LowOB...4..136G}
{Giclas}, H.~L., {Slaughter}, C.~D., \& {Burnham}, R. 1959, Lowell Observatory
  Bulletin, 4, 136

\bibitem[{{Goodman} \& {Weare}(2010)}]{2010CAMCS...5...65G}
{Goodman}, J., \& {Weare}, J. 2010, Communications in Applied Mathematics and
  Computational Science, 5, 65, \dodoi{10.2140/camcos.2010.5.65}

\bibitem[{{Gudin} {et~al.}(2021){Gudin}, {Shank}, {Beers}, {Yuan}, {Limberg},
  {Roederer}, {Placco}, {Holmbeck}, {Dietz}, {Rasmussen}, {Hansen}, {Sakari},
  {Ezzeddine}, \& {Frebel}}]{2021ApJ...908...79G}
{Gudin}, D., {Shank}, D., {Beers}, T.~C., {et~al.} 2021, \apj, 908, 79,
  \dodoi{10.3847/1538-4357/abd7ed}

\bibitem[{{Hansen} {et~al.}(2018){Hansen}, {El-Souri}, {Monaco}, {Villanova},
  {Bonifacio}, {Caffau}, \& {Sbordone}}]{2018ApJ...855...83H}
{Hansen}, C.~J., {El-Souri}, M., {Monaco}, L., {et~al.} 2018, \apj, 855, 83,
  \dodoi{10.3847/1538-4357/aa978f}

\bibitem[{Harris {et~al.}(2020)Harris, Millman, van~der Walt, Gommers,
  Virtanen, Cournapeau, Wieser, Taylor, Berg, Smith, Kern, Picus, Hoyer, van
  Kerkwijk, Brett, Haldane, del R{\'{i}}o, Wiebe, Peterson,
  G{\'{e}}rard-Marchant, Sheppard, Reddy, Weckesser, Abbasi, Gohlke, \&
  Oliphant}]{harris2020array}
Harris, C.~R., Millman, K.~J., van~der Walt, S.~J., {et~al.} 2020, Nature, 585,
  357, \dodoi{10.1038/s41586-020-2649-2}

\bibitem[{{Hawkins} {et~al.}(2015){Hawkins}, {Jofr{\'e}}, {Masseron}, \&
  {Gilmore}}]{2015MNRAS.453..758H}
{Hawkins}, K., {Jofr{\'e}}, P., {Masseron}, T., \& {Gilmore}, G. 2015, \mnras,
  453, 758, \dodoi{10.1093/mnras/stv1586}

\bibitem[{{Hayes} {et~al.}(2018){Hayes}, {Majewski}, {Shetrone},
  {Fern{\'a}ndez-Alvar}, {Allende Prieto}, {Schuster}, {Carigi}, {Cunha},
  {Smith}, {Sobeck}, {Almeida}, {Beers}, {Carrera}, {Fern{\'a}ndez-Trincado},
  {Garc{\'\i}a-Hern{\'a}ndez}, {Geisler}, {Lane}, {Lucatello}, {Matthews},
  {Minniti}, {Nitschelm}, {Tang}, {Tissera}, \& {Zamora}}]{2018ApJ...852...49H}
{Hayes}, C.~R., {Majewski}, S.~R., {Shetrone}, M., {et~al.} 2018, \apj, 852,
  49, \dodoi{10.3847/1538-4357/aa9cec}

\bibitem[{{Haywood} {et~al.}(2018){Haywood}, {Di Matteo}, {Lehnert}, {Snaith},
  {Khoperskov}, \& {G{\'o}mez}}]{2018ApJ...863..113H}
{Haywood}, M., {Di Matteo}, P., {Lehnert}, M.~D., {et~al.} 2018, \apj, 863,
  113, \dodoi{10.3847/1538-4357/aad235}

\bibitem[{{Helmi}(2020)}]{2020ARA&A..58..205H}
{Helmi}, A. 2020, \araa, 58, 205, \dodoi{10.1146/annurev-astro-032620-021917}

\bibitem[{{Helmi} {et~al.}(2018){Helmi}, {Babusiaux}, {Koppelman}, {Massari},
  {Veljanoski}, \& {Brown}}]{2018Natur.563...85H}
{Helmi}, A., {Babusiaux}, C., {Koppelman}, H.~H., {et~al.} 2018, \nat, 563, 85,
  \dodoi{10.1038/s41586-018-0625-x}

\bibitem[{{Helmi} \& {de Zeeuw}(2000)}]{2000MNRAS.319..657H}
{Helmi}, A., \& {de Zeeuw}, P.~T. 2000, \mnras, 319, 657,
  \dodoi{10.1046/j.1365-8711.2000.03895.x}

\bibitem[{{Helmi} {et~al.}(2017){Helmi}, {Veljanoski}, {Breddels}, {Tian}, \&
  {Sales}}]{2017A&A...598A..58H}
{Helmi}, A., {Veljanoski}, J., {Breddels}, M.~A., {Tian}, H., \& {Sales}, L.~V.
  2017, \aap, 598, A58, \dodoi{10.1051/0004-6361/201629990}

\bibitem[{Hunter(2007)}]{matplotlib}
Hunter, J.~D. 2007, Computing in Science Engineering, 9, 90,
  \dodoi{10.1109/MCSE.2007.55}

\bibitem[{{Ibata} {et~al.}(2019{\natexlab{a}}){Ibata}, {Bellazzini}, {Malhan},
  {Martin}, \& {Bianchini}}]{2019NatAs...3..667I}
{Ibata}, R.~A., {Bellazzini}, M., {Malhan}, K., {Martin}, N., \& {Bianchini},
  P. 2019{\natexlab{a}}, Nature Astronomy, 3, 667,
  \dodoi{10.1038/s41550-019-0751-x}

\bibitem[{{Ibata} {et~al.}(1994){Ibata}, {Gilmore}, \&
  {Irwin}}]{1994Natur.370..194I}
{Ibata}, R.~A., {Gilmore}, G., \& {Irwin}, M.~J. 1994, \nat, 370, 194,
  \dodoi{10.1038/370194a0}

\bibitem[{{Ibata} {et~al.}(2019{\natexlab{b}}){Ibata}, {Malhan}, \&
  {Martin}}]{2019ApJ...872..152I}
{Ibata}, R.~A., {Malhan}, K., \& {Martin}, N.~F. 2019{\natexlab{b}}, \apj, 872,
  152, \dodoi{10.3847/1538-4357/ab0080}

\bibitem[{{Ishigaki} {et~al.}(2013){Ishigaki}, {Aoki}, \&
  {Chiba}}]{2013ApJ...771...67I}
{Ishigaki}, M.~N., {Aoki}, W., \& {Chiba}, M. 2013, \apj, 771, 67,
  \dodoi{10.1088/0004-637X/771/1/67}

\bibitem[{{Ishigaki} {et~al.}(2012){Ishigaki}, {Chiba}, \&
  {Aoki}}]{2012ApJ...753...64I}
{Ishigaki}, M.~N., {Chiba}, M., \& {Aoki}, W. 2012, \apj, 753, 64,
  \dodoi{10.1088/0004-637X/753/1/64}

\bibitem[{{Johnson} \& {Pilachowski}(2010)}]{2010ApJ...722.1373J}
{Johnson}, C.~I., \& {Pilachowski}, C.~A. 2010, \apj, 722, 1373,
  \dodoi{10.1088/0004-637X/722/2/1373}

\bibitem[{{Johnson} \& {Soderblom}(1987)}]{1987AJ.....93..864J}
{Johnson}, D. R.~H., \& {Soderblom}, D.~R. 1987, \aj, 93, 864,
  \dodoi{10.1086/114370}

\bibitem[{{Kurucz}(2011)}]{2011CaJPh..89..417K}
{Kurucz}, R.~L. 2011, Canadian Journal of Physics, 89, 417,
  \dodoi{10.1139/p10-104}

\bibitem[{{Lawler} {et~al.}(2019){Lawler}, {Hala}, {Sneden}, {Nave}, {Wood}, \&
  {Cowan}}]{2019ApJS..241...21L}
{Lawler}, J.~E., {Hala}, {Sneden}, C., {et~al.} 2019, \apjs, 241, 21,
  \dodoi{10.3847/1538-4365/ab08ef}

\bibitem[{{Lawler} {et~al.}(2001){Lawler}, {Wickliffe}, {den Hartog}, \&
  {Sneden}}]{2001ApJ...563.1075L}
{Lawler}, J.~E., {Wickliffe}, M.~E., {den Hartog}, E.~A., \& {Sneden}, C. 2001,
  \apj, 563, 1075, \dodoi{10.1086/323407}

\bibitem[{{Li} {et~al.}(2019){Li}, {Du}, {Liu}, {Donlon}, \&
  {Newberg}}]{2019ApJ...874...74L}
{Li}, H., {Du}, C., {Liu}, S., {Donlon}, T., \& {Newberg}, H.~J. 2019, \apj,
  874, 74, \dodoi{10.3847/1538-4357/ab06f4}

\bibitem[{{Limberg} {et~al.}(2021){Limberg}, {Rossi}, {Beers}, {Perottoni},
  {P{\'e}rez-Villegas}, {Santucci}, {Abuchaim}, {Placco}, {Lee}, {Christlieb},
  {Norris}, {Bessell}, {Ryan}, {Wilhelm}, {Rhee}, \&
  {Frebel}}]{2021ApJ...907...10L}
{Limberg}, G., {Rossi}, S., {Beers}, T.~C., {et~al.} 2021, \apj, 907, 10,
  \dodoi{10.3847/1538-4357/abcb87}

\bibitem[{{Lindegren} {et~al.}(2021){Lindegren}, {Klioner}, {Hern{\'a}ndez},
  {Bombrun}, {Ramos-Lerate}, {Steidelm{\"u}ller}, {Bastian}, {Biermann}, {de
  Torres}, {Gerlach}, {Geyer}, {Hilger}, {Hobbs}, {Lammers}, {McMillan},
  {Stephenson}, {Casta{\~n}eda}, {Davidson}, {Fabricius}, {Gracia-Abril},
  {Portell}, {Rowell}, {Teyssier}, {Torra}, {Bartolom{\'e}}, {Clotet},
  {Garralda}, {Gonz{\'a}lez-Vidal}, {Torra}, {Abbas}, {Altmann}, {Anglada
  Varela}, {Balaguer-N{\'u}{\~n}ez}, {Balog}, {Barache}, {Becciani}, {Bernet},
  {Bertone}, {Bianchi}, {Bouquillon}, {Brown}, {Bucciarelli}, {Busonero},
  {Butkevich}, {Buzzi}, {Cancelliere}, {Carlucci}, {Charlot}, {Cioni},
  {Crosta}, {Crowley}, {del Peloso}, {del Pozo}, {Drimmel}, {Esquej}, {Fienga},
  {Fraile}, {Gai}, {Garcia-Reinaldos}, {Guerra}, {Hambly}, {Hauser},
  {Jan{\ss}en}, {Jordan}, {Kostrzewa-Rutkowska}, {Lattanzi}, {Liao}, {Licata},
  {Lister}, {L{\"o}ffler}, {Marchant}, {Masip}, {Mignard}, {Mints}, {Molina},
  {Mora}, {Morbidelli}, {Murphy}, {Pagani}, {Panuzzo}, {Pe{\~n}alosa Esteller},
  {Poggio}, {Re Fiorentin}, {Riva}, {Sagrist{\`a} Sell{\'e}s}, {Sanchez
  Gimenez}, {Sarasso}, {Sciacca}, {Siddiqui}, {Smart}, {Souami}, {Spagna},
  {Steele}, {Taris}, {Utrilla}, {van Reeven}, \&
  {Vecchiato}}]{2021A&A...649A...2L}
{Lindegren}, L., {Klioner}, S.~A., {Hern{\'a}ndez}, J., {et~al.} 2021, \aap,
  649, A2, \dodoi{10.1051/0004-6361/202039709}

\bibitem[{{Lynden-Bell} \& {Lynden-Bell}(1995)}]{1995MNRAS.275..429L}
{Lynden-Bell}, D., \& {Lynden-Bell}, R.~M. 1995, \mnras, 275, 429,
  \dodoi{10.1093/mnras/275.2.429}

\bibitem[{{Mackereth} {et~al.}(2019){Mackereth}, {Schiavon}, {Pfeffer},
  {Hayes}, {Bovy}, {Anguiano}, {Allende Prieto}, {Hasselquist}, {Holtzman},
  {Johnson}, {Majewski}, {O'Connell}, {Shetrone}, {Tissera}, \&
  {Fern{\'a}ndez-Trincado}}]{2019MNRAS.482.3426M}
{Mackereth}, J.~T., {Schiavon}, R.~P., {Pfeffer}, J., {et~al.} 2019, \mnras,
  482, 3426, \dodoi{10.1093/mnras/sty2955}

\bibitem[{{Majewski} {et~al.}(2003){Majewski}, {Skrutskie}, {Weinberg}, \&
  {Ostheimer}}]{2003ApJ...599.1082M}
{Majewski}, S.~R., {Skrutskie}, M.~F., {Weinberg}, M.~D., \& {Ostheimer}, J.~C.
  2003, \apj, 599, 1082, \dodoi{10.1086/379504}

\bibitem[{{Matsuno} {et~al.}(2019){Matsuno}, {Aoki}, \&
  {Suda}}]{2019ApJ...874L..35M}
{Matsuno}, T., {Aoki}, W., \& {Suda}, T. 2019, \apjl, 874, L35,
  \dodoi{10.3847/2041-8213/ab0ec0}

\bibitem[{{Meza} {et~al.}(2005){Meza}, {Navarro}, {Abadi}, \&
  {Steinmetz}}]{2005MNRAS.359...93M}
{Meza}, A., {Navarro}, J.~F., {Abadi}, M.~G., \& {Steinmetz}, M. 2005, \mnras,
  359, 93, \dodoi{10.1111/j.1365-2966.2005.08869.x}

\bibitem[{{Monty} {et~al.}(2020){Monty}, {Venn}, {Lane}, {Lokhorst}, \&
  {Yong}}]{2020MNRAS.497.1236M}
{Monty}, S., {Venn}, K.~A., {Lane}, J. M.~M., {Lokhorst}, D., \& {Yong}, D.
  2020, \mnras, 497, 1236, \dodoi{10.1093/mnras/staa1995}

\bibitem[{{Naidu} {et~al.}(2020){Naidu}, {Conroy}, {Bonaca}, {Johnson}, {Ting},
  {Caldwell}, {Zaritsky}, \& {Cargile}}]{2020ApJ...901...48N}
{Naidu}, R.~P., {Conroy}, C., {Bonaca}, A., {et~al.} 2020, \apj, 901, 48,
  \dodoi{10.3847/1538-4357/abaef4}

\bibitem[{{Navarrete} {et~al.}(2015){Navarrete}, {Chanam{\'e}}, {Ram{\'\i}rez},
  {Meza}, {Anglada-Escud{\'e}}, \& {Shkolnik}}]{2015ApJ...808..103N}
{Navarrete}, C., {Chanam{\'e}}, J., {Ram{\'\i}rez}, I., {et~al.} 2015, \apj,
  808, 103, \dodoi{10.1088/0004-637X/808/1/103}

\bibitem[{{Navarro} {et~al.}(1997){Navarro}, {Frenk}, \&
  {White}}]{1997ApJ...490..493N}
{Navarro}, J.~F., {Frenk}, C.~S., \& {White}, S. D.~M. 1997, \apj, 490, 493,
  \dodoi{10.1086/304888}

\bibitem[{{Newberg} {et~al.}(2002){Newberg}, {Yanny}, {Rockosi}, {Grebel},
  {Rix}, {Brinkmann}, {Csabai}, {Hennessy}, {Hindsley}, {Ibata}, {Ivezi{\'c}},
  {Lamb}, {Nash}, {Odenkirchen}, {Rave}, {Schneider}, {Smith}, {Stolte}, \&
  {York}}]{2002ApJ...569..245N}
{Newberg}, H.~J., {Yanny}, B., {Rockosi}, C., {et~al.} 2002, \apj, 569, 245,
  \dodoi{10.1086/338983}

\bibitem[{{Nissen} \& {Schuster}(2010)}]{2010A&A...511L..10N}
{Nissen}, P.~E., \& {Schuster}, W.~J. 2010, \aap, 511, L10,
  \dodoi{10.1051/0004-6361/200913877}

\bibitem[{{Nissen} \& {Schuster}(2011)}]{2011A&A...530A..15N}
---. 2011, \aap, 530, A15, \dodoi{10.1051/0004-6361/201116619}

\bibitem[{{Placco} {et~al.}(2021){Placco}, {Sneden}, {Roederer}, {Lawler}, {Den
  Hartog}, {Hejazi}, {Maas}, \& {Bernath}}]{2021RNAAS...5...92P}
{Placco}, V.~M., {Sneden}, C., {Roederer}, I.~U., {et~al.} 2021, Research Notes
  of the American Astronomical Society, 5, 92, \dodoi{10.3847/2515-5172/abf651}

\bibitem[{Price-Whelan {et~al.}(2020)Price-Whelan, Sipőcz, Lenz, Greco,
  Starkman, Foreman-Mackey, Lim, Oh, Koposov, \& Major}]{gala_zenodo}
Price-Whelan, A., Sipőcz, B., Lenz, D., {et~al.} 2020, adrn/gala: v1.3, v1.3,
  Zenodo, \dodoi{10.5281/zenodo.4159870}

\bibitem[{{Price-Whelan}(2017)}]{2017JOSS....2..388P}
{Price-Whelan}, A.~M. 2017, The Journal of Open Source Software, 2, 388,
  \dodoi{10.21105/joss.00388}

\bibitem[{{Ram{\'\i}rez} {et~al.}(2012){Ram{\'\i}rez}, {Mel{\'e}ndez}, \&
  {Chanam{\'e}}}]{2012ApJ...757..164R}
{Ram{\'\i}rez}, I., {Mel{\'e}ndez}, J., \& {Chanam{\'e}}, J. 2012, \apj, 757,
  164, \dodoi{10.1088/0004-637X/757/2/164}

\bibitem[{{Reddy} {et~al.}(2006){Reddy}, {Lambert}, \& {Allende
  Prieto}}]{2006MNRAS.367.1329R}
{Reddy}, B.~E., {Lambert}, D.~L., \& {Allende Prieto}, C. 2006, \mnras, 367,
  1329, \dodoi{10.1111/j.1365-2966.2006.10148.x}

\bibitem[{{Ricker} {et~al.}(2015){Ricker}, {Winn}, {Vanderspek}, {Latham},
  {Bakos}, {Bean}, {Berta-Thompson}, {Brown}, {Buchhave}, {Butler}, {Butler},
  {Chaplin}, {Charbonneau}, {Christensen-Dalsgaard}, {Clampin}, {Deming},
  {Doty}, {De Lee}, {Dressing}, {Dunham}, {Endl}, {Fressin}, {Ge}, {Henning},
  {Holman}, {Howard}, {Ida}, {Jenkins}, {Jernigan}, {Johnson}, {Kaltenegger},
  {Kawai}, {Kjeldsen}, {Laughlin}, {Levine}, {Lin}, {Lissauer}, {MacQueen},
  {Marcy}, {McCullough}, {Morton}, {Narita}, {Paegert}, {Palle}, {Pepe},
  {Pepper}, {Quirrenbach}, {Rinehart}, {Sasselov}, {Sato}, {Seager},
  {Sozzetti}, {Stassun}, {Sullivan}, {Szentgyorgyi}, {Torres}, {Udry}, \&
  {Villasenor}}]{2015JATIS...1a4003R}
{Ricker}, G.~R., {Winn}, J.~N., {Vanderspek}, R., {et~al.} 2015, Journal of
  Astronomical Telescopes, Instruments, and Systems, 1, 014003,
  \dodoi{10.1117/1.JATIS.1.1.014003}

\bibitem[{{Roederer} {et~al.}(2014){Roederer}, {Preston}, {Thompson},
  {Shectman}, {Sneden}, {Burley}, \& {Kelson}}]{2014AJ....147..136R}
{Roederer}, I.~U., {Preston}, G.~W., {Thompson}, I.~B., {et~al.} 2014, \aj,
  147, 136, \dodoi{10.1088/0004-6256/147/6/136}

\bibitem[{{Sakari} {et~al.}(2019){Sakari}, {Roederer}, {Placco}, {Beers},
  {Ezzeddine}, {Frebel}, {Hansen}, {Sneden}, {Cowan}, {Wallerstein}, {Farrell},
  {Venn}, {Matijevi{\v{c}}}, {Wyse}, {Bland-Hawthorn}, {Chiappini}, {Freeman},
  {Gibson}, {Grebel}, {Helmi}, {Kordopatis}, {Kunder}, {Navarro}, {Reid},
  {Seabroke}, {Steinmetz}, \& {Watson}}]{2019ApJ...874..148S}
{Sakari}, C.~M., {Roederer}, I.~U., {Placco}, V.~M., {et~al.} 2019, \apj, 874,
  148, \dodoi{10.3847/1538-4357/ab0c02}

\bibitem[{Schuler \& Andrews(2021)}]{spae}
Schuler, S., \& Andrews, J. 2021, {SPAE}: Stellar Parameters, Abundances, and
  Errors.
\newblock \url{https://github.com/simon-schuler/SPAE}

\bibitem[{{Schuler} {et~al.}(2011){Schuler}, {Flateau}, {Cunha}, {King},
  {Ghezzi}, \& {Smith}}]{2011ApJ...732...55S}
{Schuler}, S.~C., {Flateau}, D., {Cunha}, K., {et~al.} 2011, \apj, 732, 55,
  \dodoi{10.1088/0004-637X/732/1/55}

\bibitem[{{Schuler} {et~al.}(2015){Schuler}, {Vaz}, {Katime Santrich}, {Cunha},
  {Smith}, {King}, {Teske}, {Ghezzi}, {Howell}, \&
  {Isaacson}}]{2015ApJ...815....5S}
{Schuler}, S.~C., {Vaz}, Z.~A., {Katime Santrich}, O.~J., {et~al.} 2015, \apj,
  815, 5, \dodoi{10.1088/0004-637X/815/1/5}

\bibitem[{{Schuster} {et~al.}(2006){Schuster}, {Moitinho}, {M{\'a}rquez},
  {Parrao}, \& {Covarrubias}}]{schuster}
{Schuster}, W.~J., {Moitinho}, A., {M{\'a}rquez}, A., {Parrao}, L., \&
  {Covarrubias}, E. 2006, \aap, 445, 939, \dodoi{10.1051/0004-6361:20053796}

\bibitem[{{Schuster} {et~al.}(2019){Schuster}, {Moreno}, \&
  {Fern{\'a}ndez-Trincado}}]{2019IAUS..344..134S}
{Schuster}, W.~J., {Moreno}, E., \& {Fern{\'a}ndez-Trincado}, J.~G. 2019, in
  Dwarf Galaxies: From the Deep Universe to the Present, ed. K.~B.~W. {McQuinn}
  \& S.~{Stierwalt}, Vol. 344, 134--138, \dodoi{10.1017/S174392131800683X}

\bibitem[{{Schuster} {et~al.}(2012){Schuster}, {Moreno}, {Nissen}, \&
  {Pichardo}}]{2012A&A...538A..21S}
{Schuster}, W.~J., {Moreno}, E., {Nissen}, P.~E., \& {Pichardo}, B. 2012, \aap,
  538, A21, \dodoi{10.1051/0004-6361/201118035}

\bibitem[{{Sharma} {et~al.}(2011){Sharma}, {Bland-Hawthorn}, {Johnston}, \&
  {Binney}}]{2011ApJ...730....3S}
{Sharma}, S., {Bland-Hawthorn}, J., {Johnston}, K.~V., \& {Binney}, J. 2011,
  \apj, 730, 3, \dodoi{10.1088/0004-637X/730/1/3}

\bibitem[{{Silva} {et~al.}(2012){Silva}, {Schuster}, \&
  {Contreras}}]{2012RMxAA..48..109S}
{Silva}, J.~S., {Schuster}, W.~J., \& {Contreras}, M.~E. 2012, \rmxaa, 48, 109

\bibitem[{{Simpson} {et~al.}(2019){Simpson}, {Gargiulo}, {G{\'o}mez}, {Grand},
  {Maffione}, {Cooper}, {Deason}, {Frenk}, {Helly}, {Marinacci}, \&
  {Pakmor}}]{2019MNRAS.490L..32S}
{Simpson}, C.~M., {Gargiulo}, I., {G{\'o}mez}, F.~A., {et~al.} 2019, \mnras,
  490, L32, \dodoi{10.1093/mnrasl/slz142}

\bibitem[{{Sneden}(1973)}]{1973ApJ...184..839S}
{Sneden}, C. 1973, \apj, 184, 839, \dodoi{10.1086/152374}

\bibitem[{{Venn} {et~al.}(2004){Venn}, {Irwin}, {Shetrone}, {Tout}, {Hill}, \&
  {Tolstoy}}]{2004AJ....128.1177V}
{Venn}, K.~A., {Irwin}, M., {Shetrone}, M.~D., {et~al.} 2004, \aj, 128, 1177,
  \dodoi{10.1086/422734}

\bibitem[{{Villanova} \& {Geisler}(2011)}]{2011A&A...535A..31V}
{Villanova}, S., \& {Geisler}, D. 2011, \aap, 535, A31,
  \dodoi{10.1051/0004-6361/201117552}

\bibitem[{{Vincenzo} {et~al.}(2019){Vincenzo}, {Spitoni}, {Calura},
  {Matteucci}, {Silva Aguirre}, {Miglio}, \& {Cescutti}}]{2019MNRAS.487L..47V}
{Vincenzo}, F., {Spitoni}, E., {Calura}, F., {et~al.} 2019, \mnras, 487, L47,
  \dodoi{10.1093/mnrasl/slz070}

\bibitem[{Virtanen {et~al.}(2020)Virtanen, Gommers, Oliphant, Haberland, Reddy,
  Cournapeau, Burovski, Peterson, Weckesser, Bright, {van der Walt}, Brett,
  Wilson, Millman, Mayorov, Nelson, Jones, Kern, Larson, Carey, Polat, Feng,
  Moore, {VanderPlas}, Laxalde, Perktold, Cimrman, Henriksen, Quintero, Harris,
  Archibald, Ribeiro, Pedregosa, {van Mulbregt}, \& {SciPy 1.0
  Contributors}}]{2020SciPy-NMeth}
Virtanen, P., Gommers, R., Oliphant, T.~E., {et~al.} 2020, Nature Methods, 17,
  261, \dodoi{10.1038/s41592-019-0686-2}

\bibitem[{{Vogt} {et~al.}(1994){Vogt}, {Allen}, {Bigelow}, {Bresee}, {Brown},
  {Cantrall}, {Conrad}, {Couture}, {Delaney}, {Epps}, {Hilyard}, {Hilyard},
  {Horn}, {Jern}, {Kanto}, {Keane}, {Kibrick}, {Lewis}, {Osborne},
  {Pardeilhan}, {Pfister}, {Ricketts}, {Robinson}, {Stover}, {Tucker}, {Ward},
  \& {Wei}}]{1994SPIE.2198..362V}
{Vogt}, S.~S., {Allen}, S.~L., {Bigelow}, B.~C., {et~al.} 1994, in Society of
  Photo-Optical Instrumentation Engineers (SPIE) Conference Series, Vol. 2198,
  Instrumentation in Astronomy VIII, ed. D.~L. {Crawford} \& E.~R. {Craine},
  362, \dodoi{10.1117/12.176725}

\bibitem[{{Willman} {et~al.}(2005){Willman}, {Dalcanton}, {Martinez-Delgado},
  {West}, {Blanton}, {Hogg}, {Barentine}, {Brewington}, {Harvanek}, {Kleinman},
  {Krzesinski}, {Long}, {Neilsen}, {Nitta}, \& {Snedden}}]{2005ApJ...626L..85W}
{Willman}, B., {Dalcanton}, J.~J., {Martinez-Delgado}, D., {et~al.} 2005,
  \apjl, 626, L85, \dodoi{10.1086/431760}

\bibitem[{{Wylie-de Boer} {et~al.}(2010){Wylie-de Boer}, {Freeman}, \&
  {Williams}}]{2010AJ....139..636W}
{Wylie-de Boer}, E., {Freeman}, K., \& {Williams}, M. 2010, \aj, 139, 636,
  \dodoi{10.1088/0004-6256/139/2/636}

\bibitem[{{Yousefi} \& {Bernath}(2018)}]{2018ApJS..237....8Y}
{Yousefi}, M., \& {Bernath}, P.~F. 2018, \apjs, 237, 8,
  \dodoi{10.3847/1538-4365/aacc6a}

\bibitem[{{Yuan} {et~al.}(2020){Yuan}, {Myeong}, {Beers}, {Evans}, {Lee},
  {Banerjee}, {Gudin}, {Hattori}, {Li}, {Matsuno}, {Placco}, {Smith},
  {Whitten}, \& {Zhao}}]{2020ApJ...891...39Y}
{Yuan}, Z., {Myeong}, G.~C., {Beers}, T.~C., {et~al.} 2020, \apj, 891, 39,
  \dodoi{10.3847/1538-4357/ab6ef7}

\bibitem[{{Zhao} \& {Chen}(2021)}]{2021SCPMA..6439562Z}
{Zhao}, G., \& {Chen}, Y. 2021, Science China Physics, Mechanics, and
  Astronomy, 64, 239562, \dodoi{10.1007/s11433-020-1645-5}

\end{thebibliography}

\end{document}